\documentclass[conference]{IEEEtran}
\IEEEoverridecommandlockouts
\usepackage{cite}
\usepackage{amsmath,amssymb,amsfonts}
\usepackage{graphicx,color,overpic,psfrag}
\usepackage{xcolor}
\usepackage{algorithmic}
\usepackage{graphicx}
\usepackage{textcomp}
\usepackage{fancyhdr}
\usepackage{xcolor}
\usepackage{graphicx}
\usepackage{bm}
\usepackage{subfigure}
\usepackage{multirow}
\usepackage{booktabs}
\usepackage{caption}
\usepackage{booktabs}
\usepackage{amsmath}
\usepackage{algorithm}
\usepackage{algorithmic}
\usepackage{amsfonts}
\usepackage{xcolor}
\usepackage{tabularx}

\usepackage{booktabs,multirow}
\usepackage{graphicx}
\usepackage{subcaption}
\usepackage{siunitx}
\usepackage{colortbl}
\usepackage{enumitem}
\setlist[enumerate,1]{label=\arabic*.)}
\usepackage{cite}     
\usepackage[hidelinks]{hyperref}

\makeatletter
\renewcommand\p@paragraph{}
\makeatother

\usepackage{hyperref}
\usepackage{xcolor}
\usepackage{tabularx}
\hypersetup{
  colorlinks=true,
  linkcolor=blue,  
  citecolor=blue,  
  urlcolor=blue,   
  linkbordercolor={0 0 1},
  pdfborderstyle={/S/U/W 1} 
}

\renewcommand{\cite}[1]{\textcolor{blue}{[\citeonline{#1}]}}

\def\BibTeX{{\rm B\kern-.05em{\sc i\kern-.025em b}\kern-.08em
    T\kern-.1667em\lower.7ex\hbox{E}\kern-.125emX}}

\usepackage{color}

\begin{document}

\title{Digital Twin-Based Channel Generation Toolchain and Foundation Model for Low-Altitude XL-MIMO

\author{Mengyuan Li, \textit{Graduate Student Member, IEEE}, Yu Han, \textit{Member, IEEE}, \\ Jiachen Tian, \textit{Graduate Student Member, IEEE}, Chao-Kai Wen, \textit{Fellow, IEEE}, and Shi Jin, \textit{Fellow, IEEE}\\
}

\thanks{
This work was supported in part by the National Natural Science Foundation of China (NSFC) under Grants U25A20392, 62422105, and 624B2038, in part by the Key Technologies RD Program of Jiangsu (Prospective and Key Technologies for Industry) under Grants BE2023022, BE2023022-1, and BE2023022-2, and in part by the Southeast University Innovation Capability Enhancement Plan for Doctoral Students under Grant CXJH\_SEU 26093. The work of C.-K. Wen was supported in part by the National Science and Technology Council of Taiwan under the grant NSTC 114-2221-E-110-031-MY3.\textit{ (Corresponding authors: Yu Han; Shi Jin.)}

M. Li, Y. Han, J. Tian, and S. Jin are with the School of Information Science and Engineering, Southeast University, Nanjing 210096, China (email: mengyuan$\_$li@seu.edu.cn; hanyu@seu.edu.cn; tianjiachen@seu.edu.cn; jinshi@seu.edu.cn).

Chao-Kai Wen is with the Institute of Communications Engineering, National Sun Yat-sen University, Kaohsiung 80424, Taiwan. (e-mail: chaokai.wen@mail.nsysu.edu.tw).}}

\maketitle

\begin{abstract}
The rapid development of the low-altitude economy (LAE) has created growing demand for reliable aerial communication systems. Extremely large-scale multiple-input multiple-output (XL-MIMO) is a promising enabler for such systems due to its high spatial resolution and robust connectivity. However, three-dimensional (3D) mobility together with near-field propagation makes it difficult to obtain dedicated high-fidelity wireless datasets, hindering systematic algorithm development and evaluation. To address this issue, we develop LAETwin-XL, a digital twin (DT)-based toolchain and dataset for XL-MIMO research in LAE scenarios. Built on the Sionna ray-tracing (RT) module, the proposed toolchain simulates both near-field and far-field channels with diverse wireless labels for practical environments. Building on this dataset, we further develop a conditional denoising diffusion implicit model (CDDIM)-based generative foundation model that is pretrained to learn transferable XL-MIMO channel representations from incomplete channel observations. Unlike conventional task-specific or foundation models that rely on relatively complete channel inputs, the proposed model can generatively infer informative channel representations from partially observed channels. Experimental results demonstrate that the proposed framework achieves effective zero-shot channel extrapolation performance. Furthermore, using lightweight task heads and limited training data, it enables parameter-efficient transfer to various downstream tasks (e.g., channel estimation, classification, and localization), delivering high accuracy and robustness even under sparse antenna observations. The codes and dataset are available at \url{https://github.com/Lmyxxn/LAETwin-XL}. 
\end{abstract}


\section{Introduction}
\IEEEPARstart{L}{ow}-altitude economy (LAE) has recently attracted growing attention in both industry and academia, driven by urban air-mobility policies, increasing deployment demands, and rapid advances in electric vertical takeoff and landing (eVTOL) and unmanned aerial vehicle (UAV) technologies~\cite{LAE1,LAE2}. These advances have enabled a wide range of low-altitude applications, including aerial transportation, environmental monitoring, emergency response, and security surveillance. To support these services, reliable air-ground wireless links are indispensable for continuous positioning, telemetry, flight planning, and integrated sensing~\cite{LAE3,LAE4}. As a result, robust wireless communication and sensing technologies for LAE scenarios have become increasingly important.

Extremely large-scale multiple-input multiple-output (XL-MIMO) is widely regarded as a promising enabler for future 6G wireless communication systems, owing to its high spatial resolution and high spectral efficiency~\cite{XLMIOMO1}. With these advantages, XL-MIMO becomes well-suited to support a variety of wireless tasks, including user classification, user localization, beam management, and integrated sensing and communication. In low-altitude scenarios, the propagation environment often exhibits sparse scattering and dominant line-of-sight (LoS) components~\cite{Jiang_ISAC_LAE_2025}, leading to more pronounced spatial structures in XL-MIMO channels and more effective exploitation of array gains. Nevertheless, most existing XL-MIMO studies still focus on terrestrial scenarios, with relatively limited attention paid to low-altitude systems~\cite{Lu_Dai_NF_CE_2023, Li_KeypointNF_Localization_2025, Zhang_6G_FF_to_NF_2023,largeailae2025}.

\begin{table*}[!t]
\centering
\caption{Comparison of representative open-source measured and simulated wireless datasets and generators.}
\label{tab:dataset_comparison}
\resizebox{\textwidth}{!}{%
\begin{tabular}{@{}llcccccc@{}} 
\toprule
\textbf{Category} & \textbf{Dataset/Generator} & \textbf{BS Antenna} & \textbf{Near-Field} & \textbf{3D UE} & \textbf{Customizable} & \textbf{DT Scenario} & \textbf{Road-Guided Support} \\ 
\midrule

Measured 
& DeepSense 6G~\cite{alkhateeb2023deepsense}  & 16 ULA             & $\times$     & $\checkmark$ & $\times$     & --           & $\checkmark$ \\ 

\midrule

\multirow{5}{*}{Simulated} 
& DeepMIMO~\cite{alkhateeb2019deepmimo}       & $16 \times 16$ UPA, Custom  & $\times$     & $\times$     & $\times$     & $\times$     & $\times$     \\
& BUPTCMCC-6G~\cite{yu2025buptcmcc}           & Custom             & $\checkmark$ & $\times$     & $\checkmark$ & $\times$     & $\times$     \\
& CAVIAR6G~\cite{borges2024caviar}            & $\le 256$ elements, Custom  & $\times$    & $\checkmark$ & $\checkmark$ & $\checkmark$ & $\checkmark$ \\
& DeepVerse 6G~\cite{Jiang_Digital_2023}      & 16 ULA, Custom             & $\times$     & $\checkmark$ & $\checkmark$ & $\times$     & $\checkmark$ \\
& \textbf{LAETwin-XL}                         & \textbf{$64 \times 64$ UPA, Custom} & \textbf{$\checkmark$} & \textbf{$\checkmark$} & \textbf{$\checkmark$} & \textbf{$\checkmark$} & \textbf{$\checkmark$} \\

\bottomrule
\end{tabular}%
}
\end{table*}

Meanwhile, from the algorithmic perspective, foundation models have recently attracted increasing interest in wireless communications. With their strong generalization ability demonstrated in natural language processing, computer vision, and speech~\cite{bommasani2021foundation}, they are increasingly viewed as a promising paradigm for learning transferable representations and supporting multiple downstream wireless tasks~\cite{promotgjj2025,Liu_LLM4CP_2024,Sheng2025BeamLLM,largeailae2025,Xu_LLM_NF_LAE_2025,alikhani2024largewirelessmodellwm,LWLM}. Existing studies on foundation models for wireless communications can be broadly divided into two categories.
The first category directly adapts foundation models pretrained in other domains to wireless tasks. Representative examples include prompt-enabled channel state information (CSI) feedback~\cite{promotgjj2025}, large language model (LLM)-based channel prediction~\cite{Liu_LLM4CP_2024}, and LLM-assisted beam prediction~\cite{Sheng2025BeamLLM}. In low-altitude scenarios, recent efforts have also explored large model-based secure communications~\cite{largeailae2025} and near-field or far-field user distinction with joint precoding and power allocation~\cite{Xu_LLM_NF_LAE_2025}. Such approaches benefit from powerful cross-domain priors and transferability, but they are typically large in size and not fully specialized to the physical characteristics of wireless channels.
The second category develops wireless-specific foundation models. For example,~\cite{alikhani2024largewirelessmodellwm} proposed a self-supervised Transformer-based foundation model for massive MIMO channels, while~\cite{LWLM} designed a localization foundation model with multiple sparse-domain transformations.
Compared with directly adapting general-purpose foundation models, these wireless-specific designs are often more compact and can better exploit domain structure. However, their effectiveness critically depends on the availability of large-scale, high-fidelity wireless data, which is particularly challenging in low-altitude XL-MIMO systems.

As summarized in Table~\ref{tab:dataset_comparison}, existing open-source measured and simulated wireless datasets or generators, including DeepSense 6G~\cite{alkhateeb2023deepsense}, DeepMIMO~\cite{alkhateeb2019deepmimo}, BUPTCMCC-6G~\cite{yu2025buptcmcc}, CAVIAR6G~\cite{borges2024caviar}, and DeepVerse 6G~\cite{Jiang_Digital_2023}, still do not jointly support XL-MIMO, 3D low-altitude scenarios, and near-field propagation. This limitation motivates the development of a tailored dataset generation framework for LAE XL-MIMO systems. Collecting such data in real scenarios is costly and inflexible, especially when one needs diverse 3D trajectories, controllable environments, and per-antenna channel responses. In this context, digital twins (DTs) provide a practical alternative. By combining 3D scene reconstruction, realistic material assignment, and ray-tracing-based electromagnetic simulation, DTs can generate channel data that closely match practical propagation environments~\cite{jiang2024dtisac}. Representative DT-assisted studies include space-air-ground integrated edge computing~\cite{He2025TMC_DT_SAGIMEC}, DT-enabled fingerprint database construction~\cite{LocalizationinDigitalTwin}, and DT-assisted UAV trajectory planning in unknown environments~\cite{TrajectoryDesignLAE}. These studies demonstrate the value of DTs in wireless system design and highlight their potential for data-driven wireless research.

Building on the above observations, we develop a DT-driven XL-MIMO foundation-model framework for LAE. The framework consists of \textbf{LAETwin-XL}, a customizable DT-based dataset with near-field and 3D scene support, and a CDDIM-based foundation model that learns transferable XL-MIMO representations from incomplete observations. Built upon this pretrained backbone, the framework can be efficiently adapted to multiple downstream wireless tasks. The main contributions of this work are summarized as follows:
\begin{itemize}
\item \textbf{DT-driven low-altitude XL-MIMO dataset generation:}
We develop a high-fidelity DT-driven channel generation toolchain for low-altitude XL-MIMO systems, including 3D scene construction, realistic material assignment, road-guided UAV trajectory generation support, and per-antenna ray-tracing-based channel synthesis. Based on this toolchain, we construct and release the \textbf{LAETwin-XL} dataset, covering 13 cities under both urban and suburban scenarios, with 57,236 samples for pretraining and 13,450 samples for downstream tasks.

\item \textbf{Generative XL-MIMO foundation model under incomplete observations:}
We propose a CDDIM-based foundation model for XL-MIMO channels, pretrained through conditional generative reconstruction from partial observations. To benchmark the in-domain generalization capability of this pretrained model, we evaluate its performance on the channel extrapolation task. Unlike conventional task-specific models and existing wireless foundation models, which typically rely on complete CSI inputs, the proposed model learns transferable representations directly from incomplete observations.

\item \textbf{Parameter-efficient downstream adaptation:}
Built upon the pretrained foundation model, the proposed framework can be adapted to channel estimation, user classification, and localization using lightweight task-specific heads, limited fine-tuning, and a relatively small amount of downstream data. Compared with designing dedicated models separately for different tasks, this unified framework provides a more efficient adaptation paradigm while maintaining competitive performance across multiple downstream tasks.

\item \textbf{Comprehensive experimental validation:}
Extensive experiments validate both the utility of \textbf{LAETwin-XL} and the effectiveness of the proposed framework. The results show that, under sparse antenna activation and low signal-to-noise ratio (SNR) conditions, the proposed method remains robust across multiple downstream tasks.
\end{itemize}

\textbf{Notations.}
Scalars are denoted by regular letters, while vectors and matrices are denoted by lowercase and uppercase boldface letters, respectively. $\mathbb{R}$ and $\mathbb{C}$ denote the sets of real and complex numbers. $\mathcal{N}$ and $\mathcal{CN}$ denote the real and complex Gaussian distributions, respectively. The Euclidean norm is denoted by $\|\cdot\|$. Superscripts $(\cdot)^{\mathsf{T}}$ and $(\cdot)^{\mathsf{H}}$ represent the transpose and conjugate transpose, respectively. The expectation operator is denoted by $\mathbb{E}\{\cdot\}$. $\Re\{\cdot\}$ and $\Im\{\cdot\}$ denote the real and imaginary parts, respectively. $\mathbf{I}$ denotes the identity matrix. $\odot$ denotes the Hadamard product. The operator $\bmod$ denotes modulo reduction.

\section{System Model}
\label{sec:system_model}

In this section, we introduce the considered low-altitude XL-MIMO system and describe near-field and far-field channel models, which form the basis for subsequent dataset generation, task formulation, and algorithm design.

\subsection{Low-Altitude XL-MIMO System}

As depicted in Fig.~\ref{fig:LAE system}, we consider the uplink transmission of a low-altitude XL-MIMO system operating in both urban and suburban scenarios. In this system, each UAV serves as a single-antenna user equipment (UE) located at $\mathbf{o}_{\text{T}} \in \mathbb{R}^3$, communicating with a base station (BS) centered at $\mathbf{o}_{\text{R}} \in \mathbb{R}^3$.
The BS is equipped with a uniform planar array (UPA) consisting of $M = M_y \times M_z$ elements. Assuming the UPA is deployed in the $y$-$z$ plane and oriented along the $+x$ axis, the 3D coordinate of the $m$-th antenna element, which is indexed by $(m_y, m_z)$, can be expressed as
\begin{equation}
    \mathbf{p}_{\text{R}, m} = \mathbf{o}_\text{R} + \Big[ 0, \; \big(m_y - \tfrac{M_y-1}{2}\big)d_y, \; \big(\tfrac{M_z-1}{2} - m_z\big)d_z \Big]^{\mathsf{T}},
\end{equation}
where $d_y$ and $d_z$ denote the horizontal and vertical inter-element spacings, while $m_y \in \{0,\ldots,M_y-1\}$ and $m_z \in \{0,\ldots,M_z-1\}$ represent the corresponding antenna indices.
Given the UPA aperture $D$, the Rayleigh distance is defined as $D_{\text{R}} = 2D^2/{\lambda}$, with $\lambda$ being the carrier wavelength. Due to its mobility throughout the LAE, the UAV trajectory spans both the near-field and far-field regions.

\subsection{Near-Field and Far-Field Channel Models}

\begin{figure}[t]
    \centering
    \includegraphics[width=0.8\linewidth]{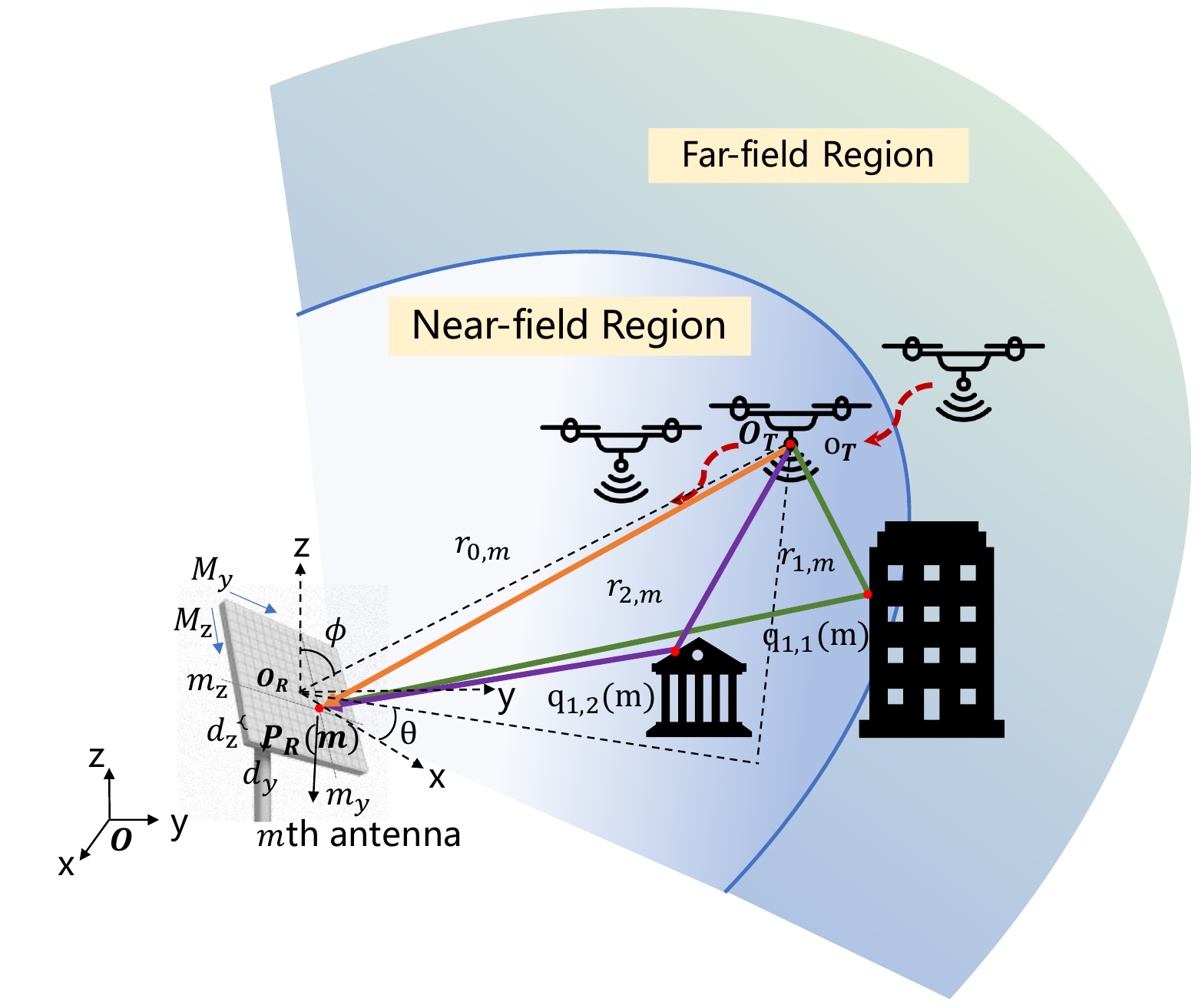}
    \caption{The considered low-altitude XL-MIMO system, featuring a UAV (UE) communicating with the ground-based large-scale UPA at the BS.}

    \label{fig:LAE system}
\end{figure}

We employ the Sionna RT module~\cite{sionnaRT} to simulate high-fidelity XL-MIMO channels. By providing exact per-antenna channel impulse response computations through a combination of shooting-and-bouncing rays, the image method, and hashing-based deduplication to eliminate redundant paths, Sionna RT inherently supports accurate near-field channel modeling, which naturally degenerates to the far-field regime beyond the Rayleigh distance $D_{\text{R}}$.
As illustrated in Fig.~\ref{fig:LAE system}, the wireless propagation between the UE and the $m$-th BS antenna comprises $L_m$ propagation paths, indexed by $l \in \{1, \dots, L_m\}$. The geometric length of each path depends on its specific propagation path. For the LoS path ($l=1$), the distance between the UE at $\mathbf{o}_{\text{T}}$ and the $m$-th BS element at $\mathbf{p}_{\text{R}, m}$ is
\begin{equation}
    r_{1,m} = \|\mathbf{o}_{\text{T}} - \mathbf{p}_{\text{R}, m}\|.
\end{equation}
For the $l$-th non-line-of-sight (NLoS) path ($l > 1$) undergoing $K_l$ environmental interactions, the total path length aggregates the Euclidean distances between consecutive scatterers as:
\begin{equation}
\resizebox{0.89\columnwidth}{!}{$
    r_{l,m} = \|\mathbf{o}_{\text{T}} - \mathbf{q}_{l,1}^{(m)}\|+ \sum_{k=1}^{K_l-1} \|\mathbf{q}_{l,k}^{(m)} - \mathbf{q}_{l,k+1}^{(m)}\| + \|\mathbf{q}_{l,K_l}^{(m)} - \mathbf{p}_{\text{R}, m}\|,
$}
\end{equation}
where $\mathbf{q}_{l,k}^{(m)}$ denotes the 3D coordinate of the $k$-th scattering point corresponding to the ray reaching the $m$-th antenna on the $l$-th path. 

Based on these geometric traits, Sionna RT simulates a set of $L_m$-path parameters for the $m$-th receive antenna, which is denoted as
\begin{equation}
    \mathcal{P}_m = \Big\{\, r_{l,m},\, g_{l,m},\, \big\{ \boldsymbol{\omega}_{l,k}^{(m)} \big\}_{k=1}^{K_l} \,\Big\}_{l=1}^{L_m},
\end{equation}
where $g_{l,m}$ represents the complex path gain. The term $\boldsymbol{\omega}_{l,k}^{(m)} = \big(\theta^{\mathrm{R}, (m)}_{l,k}, \phi^{\mathrm{R}, (m)}_{l,k}, \theta^{\mathrm{T}}_{l,k}, \phi^{\mathrm{T}}_{l,k}\big)$ encapsulates the angular information for the $k$-th interaction, with $(\theta^{\mathrm{R}}, \phi^{\mathrm{R}})$ and $(\theta^{\mathrm{T}}, \phi^{\mathrm{T}})$ being the azimuth and elevation angles of arrival (AoA) at the receiver and departure (AoD) at the transmitter, respectively. Following the front-facing UPA geometry, we use $\theta\in[-\pi/2,\pi/2]$ to denote the azimuth angle within the visible half-space and $\phi\in[0,\pi]$ to denote the elevation angle. For LoS paths, the AoDs and AoAs are directly determined by the geometric link between the UE and the BS antennas. For NLoS paths, they are defined by the directions toward the first and last interaction points, respectively.

\begin{figure*}[t]
    \centering
    \includegraphics[width=0.95\linewidth]{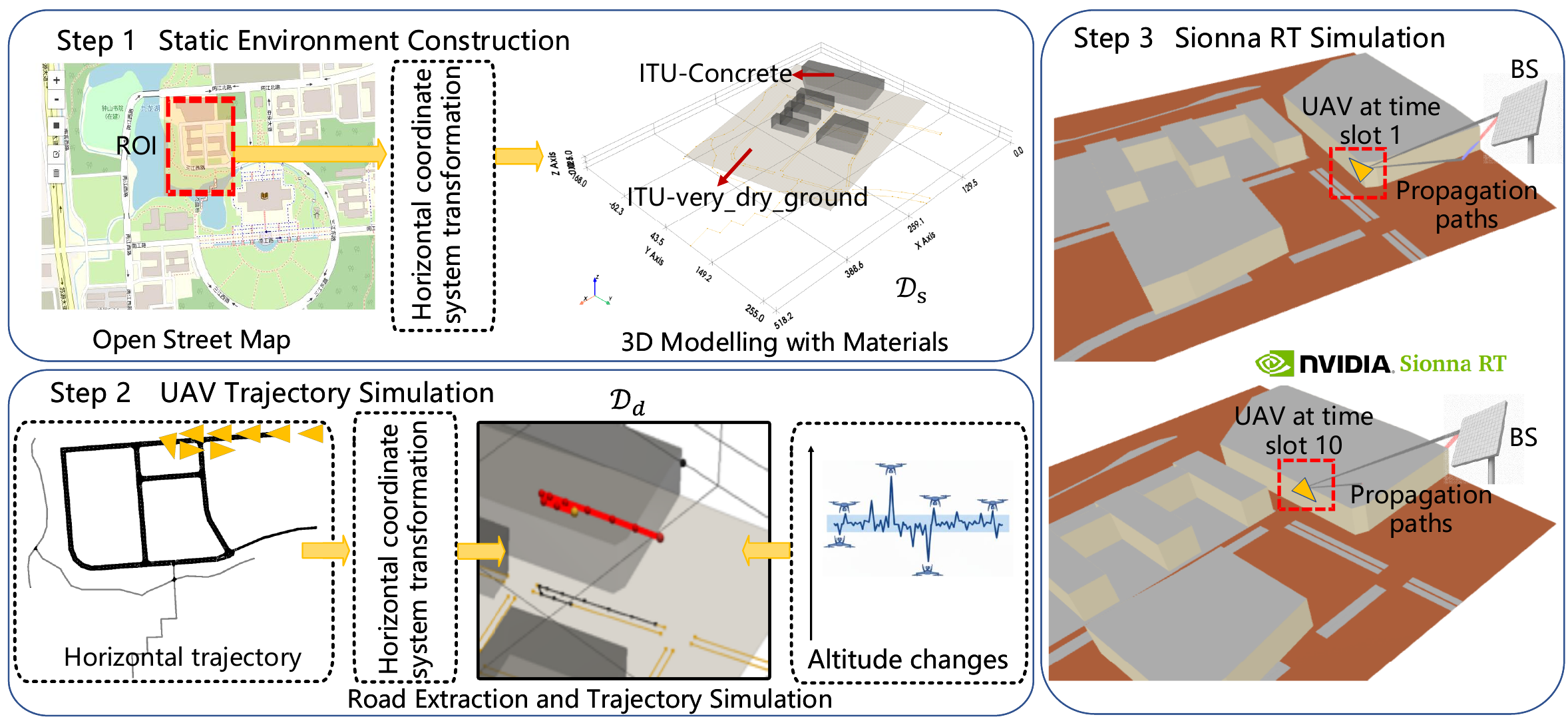}
    \caption{The overall workflow of the proposed channel generation toolchain for low-altitude XL-MIMO systems.}
    \label{fig:dataset construction}
\end{figure*}

The near-field channel between the moving UE and the $m$-th BS antenna element at subcarrier frequency $f$ and observation time $\tau$ is modeled as:
\begin{equation}
    h_{m}(f,\tau) = \sum_{l=1}^{L_m} g_{l,m} \, e^{-j \frac{2\pi f}{c} r_{l,m}} \, e^{j 2\pi \nu_{l,m} \tau},
    \label{eq:channel_model}
\end{equation}
where $\nu_{l,m}$ denotes the Doppler shift of the $l$-th path observed at the $m$-th antenna, $c$ represents the speed of light. Since the BS array and environmental scatterers are static in the Sionna RT setup, the Doppler shift $\nu_{l,m}$ is solely induced by the UE's movement. $\tau = n T_{\text{sym}}$ denotes the observation of the $n$-th OFDM symbol, where $T_{\text{sym}}$ is the duration of an OFDM symbol. Let $\mathbf{v}_{\tau} \in \mathbb{R}^{3\times 1}$ denote the 3D physical velocity vector of the UE at time $\tau$. The Doppler shift is determined by projecting $\mathbf{v}_{\tau}$ onto the signal departure direction:
\begin{equation}
    \nu_{l,m} = \frac{1}{\lambda}\mathbf{v}_{\tau}^{\mathsf{T}} \hat{\mathbf{u}}_{l,m},
    \label{eq:doppler_near}
\end{equation}
where $\hat{\mathbf{u}}_{l,m}$ is the unit departure-direction vector from the UE for the $l$-th path.
The complex gain $g_{l,m}$ is given by
\begin{equation}
\label{eq:complex_gain}
\begin{split}
    g_{l,m}
= \frac{\lambda}{4\pi r_{l,m}}\,
\mathbf c_{\mathrm R}^{\mathsf H}(\Omega_{\mathrm R,l,m})
\left(\prod_{k=1}^{K_l}\mathbf{T}_{l,k}^{(m)}\right)
\mathbf c_{\mathrm T}(\Omega_{\mathrm T,l}),
\end{split}
\end{equation}
where $\mathbf{c}_{\mathrm R}(\cdot)\in\mathbb{C}^{2\times1}$ and $\mathbf{c}_{\mathrm T}(\cdot)\in\mathbb{C}^{2\times1}$ denote the complex polarization field-pattern vectors of the receive and transmit antennas, respectively, and $\mathbf{T}_{l,k}^{(m)}\in\mathbb{C}^{2\times2}$ denotes the electromagnetic transfer matrix associated with the $k$-th interaction. Here, $\Omega_{\mathrm R,l,m}$ and $\Omega_{\mathrm T,l}$ denote the AoA and AoD associated with the corresponding path endpoints extracted from $\boldsymbol{\omega}_{l,k}^{(m)}$. By evaluating $r_{l,m}$, $\nu_{l,m}$, and $g_{l,m}$ on a per-antenna basis, the proposed model preserves the spherical-wave behavior and element-wise spatial non-stationarity of the near-field regime.

Moreover, we briefly show how the considered near-field model reduces to the conventional far-field model beyond the Rayleigh distance, thereby revealing the distinct propagation characteristics in the near-field and far-field regions. In the far-field regime, the spherical wavefront degenerates into a plane wave, and all antennas approximately observe the same set of multipath components, i.e., $L_m=L,\forall m$. 
Let $\mathbf d_m=\mathbf p_{\mathrm R,m}-\mathbf o_{\mathrm R}$ denote the relative position of the $m$-th antenna with respect to the array center. In the far-field regime, the path distance can be approximated as
\begin{equation}
    r_{l,m} \approx r_{l,0} - \mathbf u_{\mathrm R,l}^{\mathsf T}\mathbf d_m .
\end{equation}
Accordingly, the channel reduces to
\begin{equation}
    h_m(f,\tau) \approx \sum_{l=1}^{L} \bar g_l
    e^{j\frac{2\pi f}{c}\mathbf u_{\mathrm R,l}^{\mathsf T}\mathbf d_m}
    e^{j2\pi \nu_l \tau},
\label{eq:far_channel}
\end{equation}
where $\bar g_l$ absorbs the antenna-independent path gain and common phase terms. Both $\bar g_l$ and the Doppler shift $\nu_l$ become antenna-independent under the plane-wave assumption.

\section{Digital-Twin Channel Generation Toolchain}
\label{sec:dataset}

In this section, we develop a DT-based channel generation toolchain that simulates channels from virtual electromagnetic environments for XL-MIMO-assisted LAE research. The overall workflow of the toolchain is illustrated in Fig.~\ref{fig:dataset construction}, including three steps: static environment construction, UAV trajectory simulation, and Sionna RT simulation.

\subsection{3D Static Environment Construction}

\begin{figure}[t]
    \centering
    \includegraphics[width=0.99\linewidth]{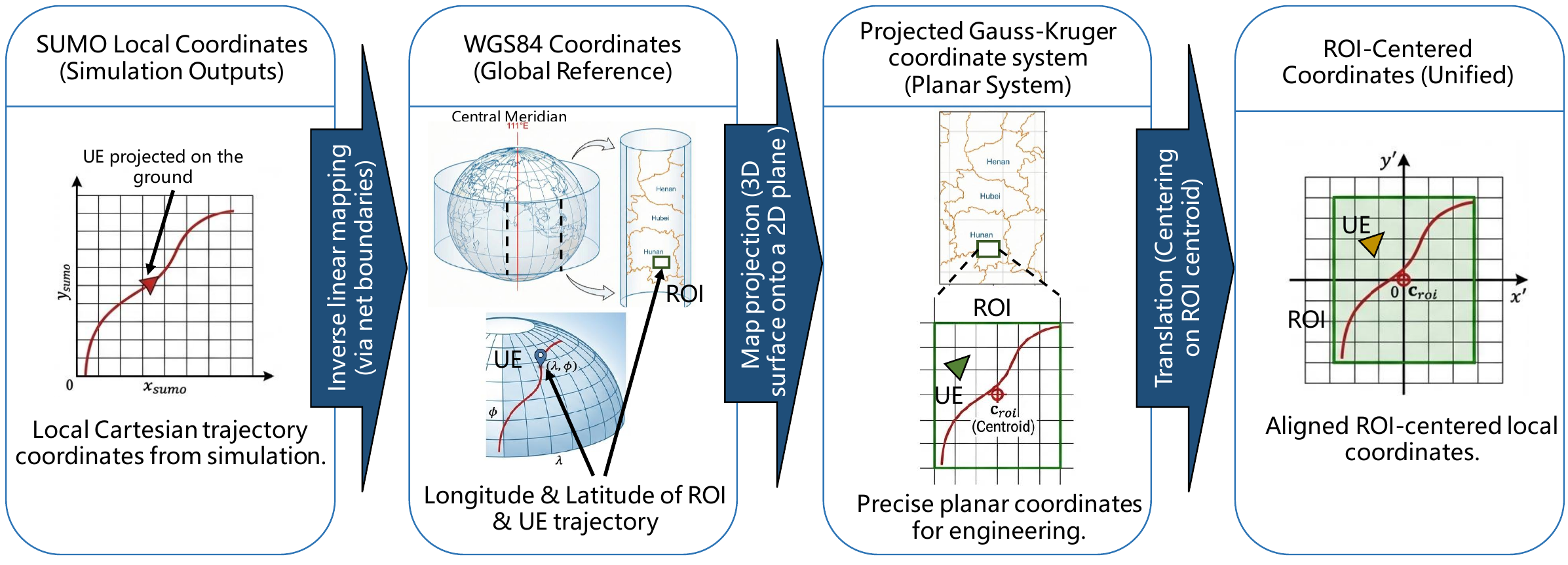}
    \caption{Horizontal coordinate system transformation of SUMO trajectories and WGS84 scenes.}
    \label{fig:coord_transform}
\end{figure}

To enable realistic simulation, we first construct the static environment $\mathcal{D}_s$ based on real-world 2D maps. In the proposed toolchain, we select the region of interest (ROI) from OpenStreetMap (OSM) to extract the geometric information of static objects such as buildings and roads. The OSM provides the geographic coordinates of the scene based on the spherical World Geodetic System 1984 (WGS84). As illustrated in Fig.~\ref{fig:coord_transform}, to unify these representations and the following UAV trajectories, we implement a horizontal coordinate transformation pipeline. For the transformation of the established static environment $\mathcal{D}_s$, we first project these spherical WGS84 coordinates into the corresponding regional Gauss-Kruger planar rectangular coordinate system (e.g., \textit{EPSG:4527} for Beijing, based on CGCS2000). Then, a global translation is applied to center the ROI at its centroid, establishing a unified metric coordinate system for each scenario. Coupled with building heights from OSM when available, and user-defined heights otherwise, we can reconstruct 3D environments with accurate geometry and consistent physical scales.

These reconstructed geometries are further registered with material properties specified by the International Telecommunication Union (ITU) recommendations to form the 3D scene model with materials. Specifically, we map surface textures to the standard material keys supported by the Sionna RT engine, which encompasses a set of ITU-compliant materials including \texttt{itu-concrete}, \texttt{itu-metal}, \texttt{itu-medium-dry-ground}, and \texttt{itu-glass}, among others~\cite{itu_p2040_3}. Each material is associated with electromagnetic attributes such as permittivity and conductivity, which affect wave propagation, reflection, and transmission in the environment. By accurately modeling both the geometry and the material composition of the wireless communication scenario, $\mathcal{D}_s$ provides a realistic and physically consistent environment for incorporating user mobility.

\subsection{UAV Trajectory Simulation}

We simulate 3D UAV trajectories across two typical operational scenarios: urban operations and suburban cruising~\cite{3gpp36777, mozaffari2017mobile,zeng2019accessing}. 
For the urban case, UAVs are assumed to fly above city roads to perform common urban tasks such as vehicle tracking and road inspection. We employ the Simulation of Urban Mobility (SUMO) framework to extract paths from OSM and generate road-guided horizontal trajectories. By setting SUMO parameters such as \texttt{period} and \texttt{maxSpeed}, we can generate the desired traffic-constrained mobility pattern. For the altitude changes, the initial altitude is uniformly sampled from $[z_{\min}, z_{\max}]$. Subsequent altitude variations at each sampling interval $\Delta{\tau}$ follow either a bounded random walk or a sinusoidal pattern. 
For the suburban case, the horizontal trajectories are generated by applying random perturbations to an initial velocity vector to simulate nearly linear flight paths, while the altitude variations follow the same two modes as in the urban scenario.
Subsequently, these mobile UAVs are integrated into the static environment $\mathcal{D}_s$ to formulate the 3D dynamic scene $\mathcal{D}_d$, which encompasses multiple urban and suburban regions extracted from OSM. A prerequisite for this integration is spatial alignment. Since SUMO outputs mobility data in the local Cartesian frame, we first inverse-map the SUMO local trajectories back to the global WGS84 reference. The subsequent transformation undergoes the same ROI-centered alignment pipeline as shown in Fig.~\ref{fig:coord_transform}, ensuring scene-trajectory co-registration.

\subsection{Sionna RT Simulation}

We import the constructed 3D environments into Sionna RT, anchoring the receiver UPA at $\mathbf{o}_{\text{R}}$ and extracting UAV positions from the generated trajectories. To capture XL-MIMO channel characteristics with sufficient accuracy, we adapt the ray tracing (RT) configuration based on the Rayleigh distance $D_{\text{R}}$. Specifically, we disable the synthetic array approximation (\texttt{synthetic\_array=False}) for near-field links to ensure precise per-element channel evaluation, while enabling it (\texttt{synthetic\_array=True}) in the far field for computational efficiency. By simulating the propagation path parameters $\mathcal{P}_{m}$ for each antenna $m$, the corresponding near-field or far-field channels can be constructed via~\eqref{eq:channel_model} or~\eqref{eq:far_channel}. Ultimately, this toolchain outputs site-specific XL-MIMO channels, time-stamped UAV coordinates, and corresponding near-field and far-field labels, providing a comprehensive wireless benchmark for low-altitude XL-MIMO systems.

\section{Large XL-MIMO LAE Foundation Model}
\label{sec:pretrain}

In this section, we introduce the proposed CDDIM-based generative foundation model for learning transferable XL-MIMO channel representations from incomplete observations. We then describe its three main components: the forward diffusion process, the noise prediction network design, and the reverse sampling process.

\subsection{Motivation and Basic Principle}

\begin{figure*}[!t]
    \centering
    \includegraphics[width=0.95\linewidth]{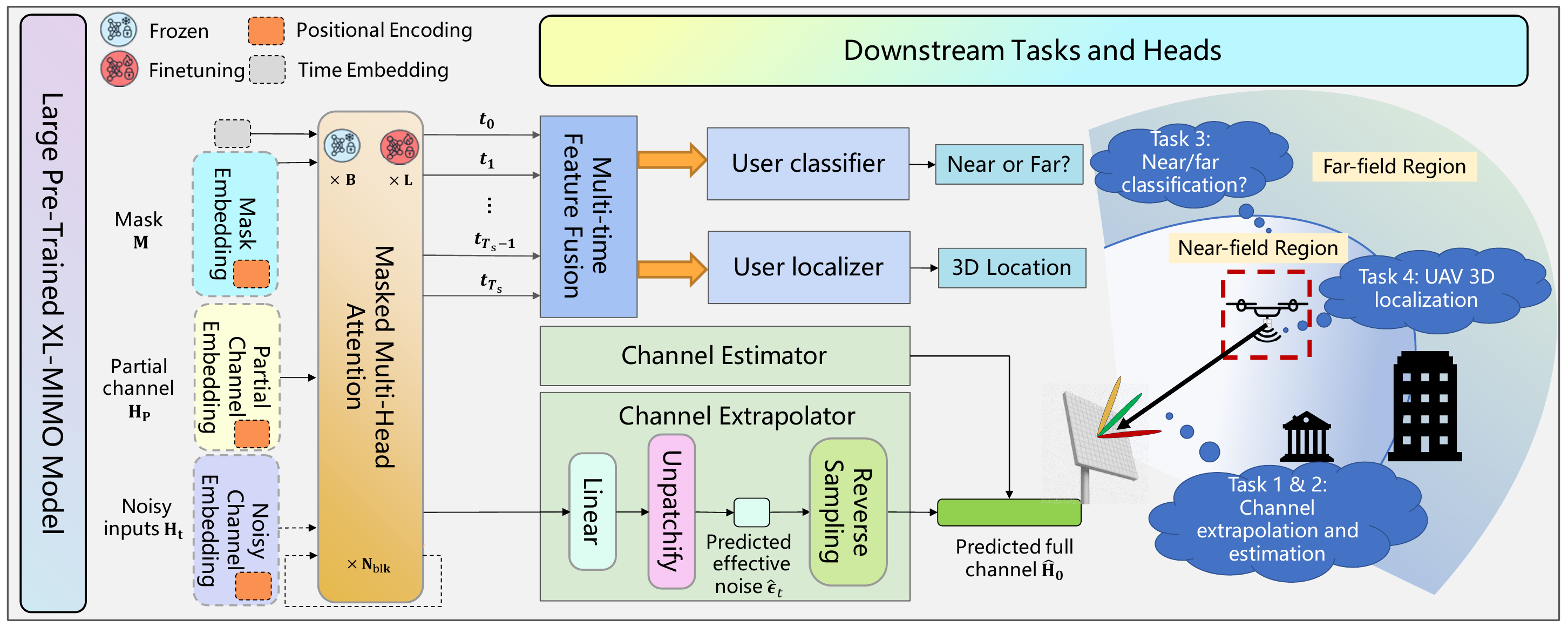}
    \caption{Overview of the proposed large generative foundation model with downstream fine-tuning. The model is pretrained on the XL-MIMO LAE dataset to learn universal channel representations, which are shared across multiple task-specific heads. During fine-tuning, the first $B$ Transformer blocks are frozen and only the last $L$ blocks are updated.
} 
    \label{fig:Large XL-MIMO Model}
\end{figure*}

In practical XL-MIMO deployments, acquiring full-dimensional CSI requires activating a large number of antennas, resulting in prohibitive hardware power consumption and measurement overhead. A promising solution is antenna-domain subsampling, which activates only a sparse subset of antennas to obtain partial channel observations. However, recovering the full channel from such incomplete observations is highly ill-posed, especially under high mask ratios.
To address this challenge, we propose a generative foundation model trained on the proposed LAETwin-XL dataset to learn transferable channel representations for XL-MIMO LAE systems. As shown in Fig.~\ref{fig:Large XL-MIMO Model}, the pretrained model serves as a unified backbone that can be efficiently adapted to diverse downstream wireless tasks through lightweight fine-tuning.

Although BERT-style masked modeling~\cite{devlin2019bert} is a widely used self-supervised paradigm in large pre-trained models, its direct application to XL-MIMO channel extrapolation remains nontrivial. Low masking ratios, e.g., 15\%, may lead to weak pretext tasks and shallow representations. Moreover, recent wireless foundation-model studies have complemented masked reconstruction with contrastive learning to enhance the structural and discriminative capability of channel representations~\cite{guler2025multitaskfoundationmodelwireless}. Inspired by this observation, we further formulate severe-mask channel extrapolation as a conditional generative task rather than relying solely on point-wise reconstruction. Specifically, we adopt CDDIM~\cite{lee2024cddim} to learn the channel distribution conditioned on the visible antenna subset and spatial coordinates, enabling more faithful reconstruction of missing channel structures.
To exploit the horizontal and vertical spatial correlations of the antenna array, we reshape the Sionna RT generated channel vector $\mathbf{h}\in \mathbb{C}^{M}$ into a 2D matrix $\mathbf{H}_0 \in \mathbb{C}^{M_y \times M_z}$ according to the UPA geometry, where $M=M_yM_z$. Based on this representation, the proposed CDDIM-based framework consists of forward diffusion, noise prediction, and reverse sampling, as detailed next.

\subsection{Forward Diffusion Process (Noise Injection)} 
Given the clean full channel $\mathbf{H}_{0}$, we define a binary mask $\mathbf{M}\in \{0,1\}^{M_y \times M_z}$, where $1$ denotes an active antenna and $0$ denotes a masked (inactive) one. The proportion of zeros in $\mathbf{M}$ is defined as the mask ratio $\rho_M$, which relates to the physical antenna selection ratio $\rho_\text{s}$ via $\rho_\text{s} = 1 - \rho_M$. The corresponding partially observed channel is then given by $\mathbf{H}_\text{p} = \mathbf{M}\odot \mathbf{H}_{0}$.
The forward diffusion process perturbs $\mathbf{H}_0$ according to a variance-preserving noise schedule. 
At any diffusion step $t\in\{1,\ldots,T\}$, the noisy channel can be sampled directly as
\begin{equation}
    \mathbf{H}_t
    =
    \sqrt{\bar{\alpha}_t}\,\mathbf{H}_0
    +
    \sqrt{1-\bar{\alpha}_t}\,\boldsymbol{\epsilon},
    \label{eq:noisy_channel}
\end{equation}
where $\boldsymbol{\epsilon} \sim \mathcal{CN}(\mathbf{0}, \mathbf{I})$ denotes the Gaussian noise, $\bar{\alpha}_t = \prod_{i=1}^t \alpha_i$ is the cumulative product of the noise-schedule parameters with $\alpha_i = 1 - \beta_i$, and $\beta_t$ denotes the variance of the added noise at step $t$ ($0 < \beta_1 < \cdots < \beta_T$).

\subsection{Noise Prediction Network}

XL-MIMO channels exhibit pronounced position-dependent pathloss and phase variations across the large-scale array. In particular, near-field propagation is characterized by spherical wavefronts, where the channel response is determined by the exact Euclidean distance from the user or scatterer to each individual antenna element, as illustrated in Sec.~\ref{sec:system_model}. This induces non-linear and non-shift-invariant spatial responses. Furthermore, in LAE scenarios, the channel characteristics heavily depend on the operating altitude and environment, ranging from sparse multipath at high altitudes to denser scattering in urban areas. These physically-grounded properties fundamentally differ from generic 2D images. Consequently, conventional convolutional architectures relying on translation invariance and localized receptive fields are inherently insufficient to capture such global, position-dependent variations across the extremely large aperture.

To effectively address these challenges, we adopt a Transformer backbone composed of $N_{\text{blk}}$ masked multi-head cross-attention (MHA) blocks, as shown in Fig.~\ref{fig:Large XL-MIMO Model}, incorporating two physically motivated designs. First, a masked cross-attention mechanism enables each queried antenna token to selectively aggregate information from valid observed antennas over the entire array. Specifically, the noisy channel $\mathbf{H}_{t}$, the partial observation $\mathbf{H}_\text{p}$, and the mask $\mathbf{M}$ are projected into latent embeddings to form the Query ($\mathbf{Q}$) and Key/Value ($\mathbf{K}, \mathbf{V}$) sequences, respectively. Since the complex-valued channel is represented by its real and imaginary components before being fed into the neural network, the latent embeddings are real-valued. The attention score matrix is computed as
\begin{equation}
    \mathbf{A} = \operatorname{Softmax}\left( \frac{\mathbf{Q}\mathbf{K}^{\mathsf{T}}}{\sqrt{d_k}} + \mathcal{M}(\mathbf{M}) \right),
\end{equation}
where $d_k$ denotes the key dimension, and $\mathcal{M}(\mathbf{M})$ is an additive mask setting entries to $0$ for active antennas and $-\infty$ for masked antennas. This prevents spurious interactions with zero-padded unobserved entries, allowing the model to adaptively recover missing components via data-driven global spatial correlations. Second, to ensure these learned correlations are informed by the actual array geometry, we inject time embeddings (encoding the diffusion step $t$ via standard sinusoidal functions~\cite{ho2020ddpm}) and 2D positional encodings (concatenating the sinusoidal embeddings of the horizontal $y$ and vertical $z$ indices). This encodes the UPA topology in the latent feature space, ensuring the near-field channel response aligns with the exact physical coordinates of each antenna element. The training objective is to minimize the mean squared error (MSE) between the ground truth noise and the predicted noise, formulated as
\begin{equation}
\mathcal{L}
= \mathbb{E}_{\mathbf{H}_{0},\,\mathbf{M},\,t,\,\boldsymbol{\epsilon}}
\Big[
\|\boldsymbol{\epsilon} - \hat{\boldsymbol{\epsilon}}_{\theta}(\mathbf{H}_{t}, \mathbf{H}_\text{p}, \mathbf{M}, t)\|_{2}^{2}
\Big].
\label{eq:extrapolation_loss}
\end{equation}

\subsection{Reverse Sampling Process}
Given the trained noise-prediction network, the CDDIM reverse process performs deterministic updates from $t=T$ down to $t=1$ with step size $\Delta_t$, conditioned on the partial observation $(\mathbf{H}_\text{p},\mathbf{M})$. 
Unlike denoising diffusion probabilistic models (DDPMs), which relies on stochastic sampling, CDDIM defines a non-Markovian sampling process that leads to a deterministic generative trajectory.
At each reverse step $t$, given the noisy channel $\mathbf{H}_{t}$ and the predicted noise $\hat{\boldsymbol{\epsilon}}_{t} = \boldsymbol{\epsilon}_{\theta}(\mathbf{H}_{t},\mathbf{H}_\text{p},\mathbf{M},t)$, the denoised channel estimate $\hat{\mathbf{H}}_{t-\Delta_t}$ at the previous time step $t-\Delta_t$ is computed as
\begin{equation}
    \hat{\mathbf{H}}_{t-\Delta_t}
    = \sqrt{\bar{\alpha}_{t-\Delta_t}} \left( \frac{\mathbf{H}_t - \sqrt{1-\bar{\alpha}_t}\hat{\boldsymbol{\epsilon}}_t}{\sqrt{\bar{\alpha}_t}} \right)
    + \sqrt{1-\bar{\alpha}_{t-\Delta_t}} \hat{\boldsymbol{\epsilon}}_t.
\label{eq:reverse_sampling}
\end{equation}
By iteratively applying~\eqref{eq:reverse_sampling}, the model progressively reconstructs the estimate of the clean full channel $\hat{\mathbf{H}}_0$ from a Gaussian noise initialization conditioned on the partial observation $\mathbf{H}_{\mathrm p}$ and the mask $\mathbf{M}$.

Building on this scheme, we conduct large-scale pretraining on the proposed LAETwin-XL dataset. The resulting foundation model learns transferable channel representations across different geographic regions and propagation regimes, and naturally supports channel extrapolation under a wide range of mask ratios. Based on this pretrained backbone, we next develop a downstream fine-tuning paradigm for multiple wireless tasks.

\section{Evaluation and Downstream Adaptation Paradigm}
\label{sec:downstream_fine-tune}

\begin{figure}[!t]
    \centering
    \includegraphics[width=0.9\linewidth]{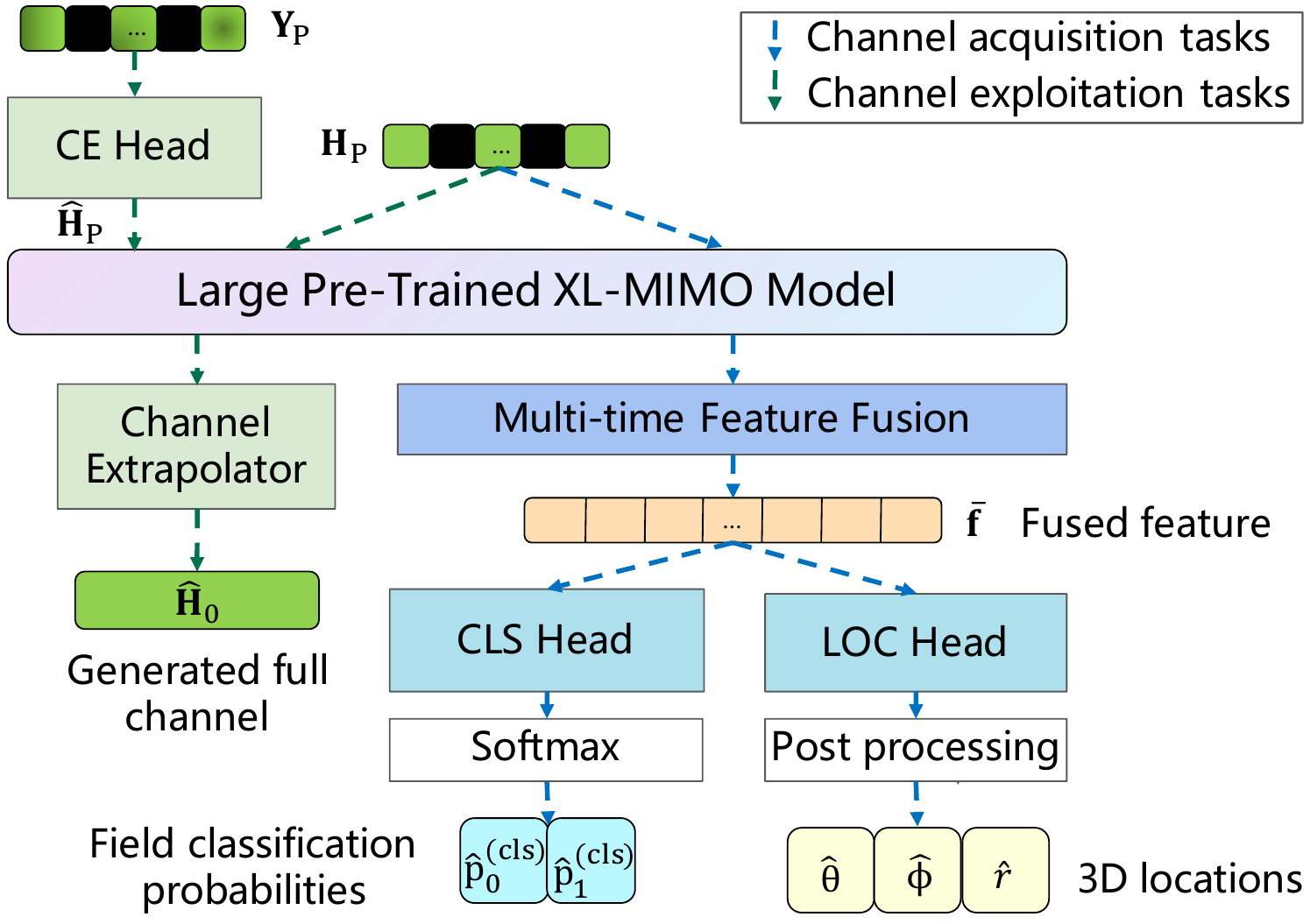}
    \caption{Evaluation and downstream adaptation paradigm. The proposed tasks are grouped into two families: channel acquisition (encompassing zero-shot generalization and fine-tuning) and channel exploitation.}
    \label{fig:downstream_tasks}
\end{figure}

As illustrated in Fig.~\ref{fig:downstream_tasks}, we evaluate the proposed foundation model from two perspectives: its in-domain generalization capability on the channel extrapolation task and its adaptability to various downstream tasks. For downstream adaptation, we attach lightweight task-specific heads to the foundation model and fine-tune the resulting model. By leveraging the channel representations learned during pretraining, this design enables efficient adaptation to different tasks. Our case study considers four representative tasks spanning channel acquisition and channel exploitation.

\subsection{Channel Acquisition Tasks}

Channel acquisition aims to reconstruct the full-dimensional XL-MIMO channel from sparse measurements. To evaluate the foundation model's generative prior, we formulate two progressive tasks. First, channel extrapolation from clean partial observations serves as an in-domain generalization benchmark to directly assess the zero-shot capability of the pretrained backbone. Second, channel estimation from noisy received pilots acts as a downstream fine-tuning task, addressing the practical challenge of joint denoising and extrapolation.

\subsubsection{Task 1: Channel extrapolation from clean partial channels}

As the in-domain generalization benchmark, this task evaluates the fundamental antenna-domain extrapolation capability of the foundation model. Since the partial observations are clean, the pretrained model directly performs zero-shot channel extrapolation without requiring any additional task-specific head. As illustrated in Fig.~\ref{fig:Large XL-MIMO Model}, conditioned on the clean partial channel $\mathbf{H}_\text{p}$, the model takes $(\mathbf{H}_t,\mathbf{H}_\text{p},\mathbf{M},t)$ as input and predicts the diffusion noise $\hat{\boldsymbol{\epsilon}}_t$ at each time step through the built-in linear projection and unpatchify modules. The full channel $\hat{\mathbf{H}}_0$ is then progressively reconstructed from $t=T$ to $t=0$ via the CDDIM reverse sampling process in~\eqref{eq:reverse_sampling}. This generative mechanism enables accurate channel reconstruction over a wide range of mask ratios $\rho_M$.

\subsubsection{Task 2: Channel estimation from received noisy pilots}

For the channel estimation task, we consider a partial-antenna selection scheme, where pilots are received only at a subset of antennas, and the goal is to recover the full XL-MIMO channel from these partial noisy observations. The received signal is modeled as
\begin{equation} 
  \mathbf{y}_\text{p} = \mathbf{S}\mathbf{h} + \mathbf{n}, 
  \quad \mathbf{y}_\text{p} \in \mathbb{C}^{P \times 1},
\end{equation}
where $\mathbf{S} \in \{0,1\}^{P \times M}$ is the antenna selection matrix that selects $P$ ($P<M$) antennas for pilot reception, and $\mathbf{n} \sim \mathcal{CN}(\mathbf{0}, \sigma^2 \mathbf{I}) \in \mathbb{C}^{P \times 1}$ denotes additive white Gaussian noise. The antenna selection ratio is defined as $\rho_s = P/M$. To align with the input format of the foundation model, we first construct a zero-padded vector $\tilde{\mathbf{y}}_\text{p} \in \mathbb{C}^{M\times 1}$ by setting the entries corresponding to unobserved antennas to zero, and then reshape it into a grid $\mathbf{Y}_\text{p} \in \mathbb{C}^{M_y\times M_z}$. Let $\mathbf{Y} \in \mathbb{C}^{M_y\times M_z}$ denote the received pilot when all antennas are activated. Then, the partial observation can be written as
\begin{equation}
  \mathbf{Y}_\text{p} = \mathbf{Y} \odot \mathbf{M}.
\end{equation}

\begin{figure}[!t]
    \centering
    \includegraphics[width=0.98\linewidth]{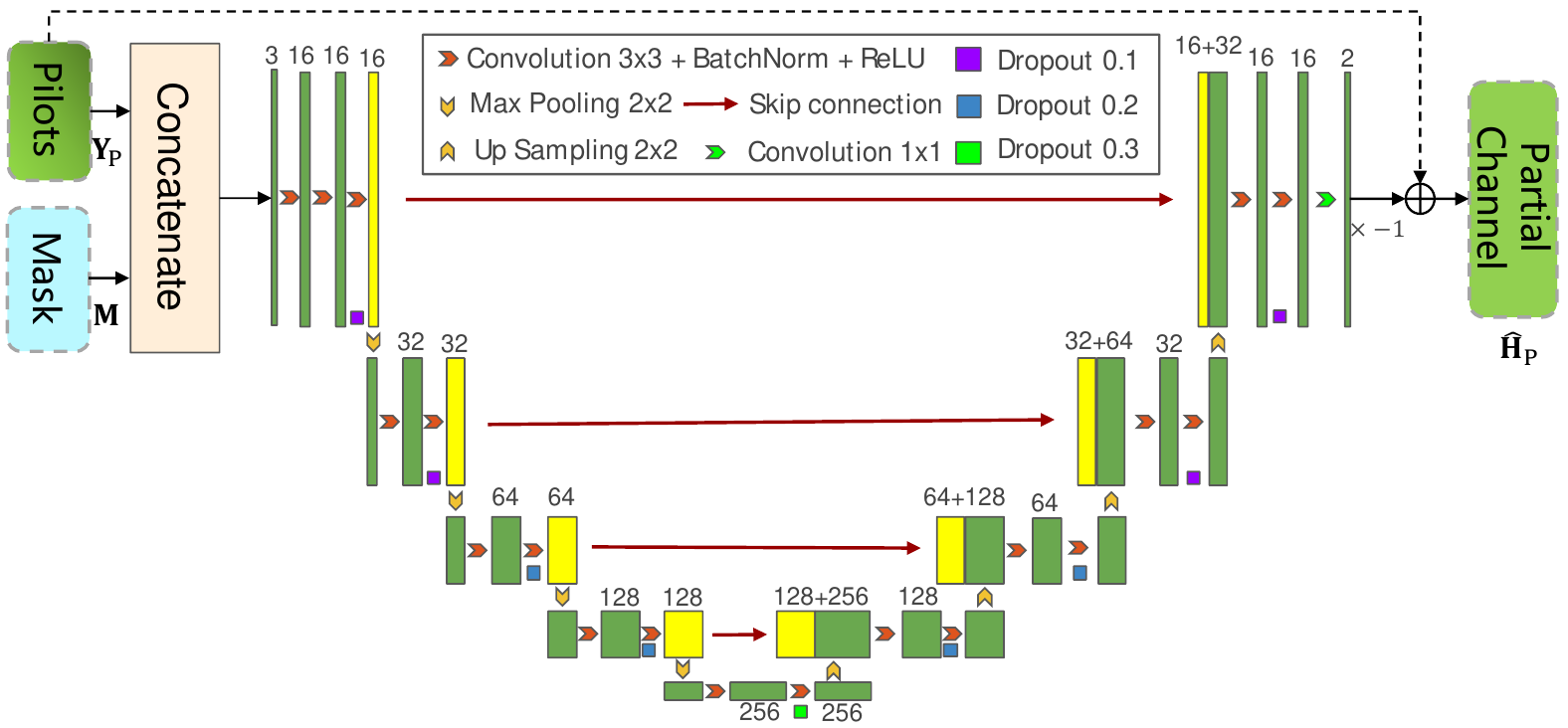}
    \caption{U-Net architecture-based denoising head for channel estimation, consisting of four downsampling stages and four upsampling stages.}
    \label{fig:CE_head}
\end{figure}

As illustrated in Fig.~\ref{fig:CE_head}, to mitigate the noise effect before channel extrapolation, we introduce a U-Net-based channel estimation (CE) head $\mathcal{U}(\cdot)$ for denoising prior to the foundation model. To preserve consistency with the observed pilots, the denoised partial channel is obtained as

\begin{equation}
  \hat{\mathbf{H}}_\text{p} = \left( \mathbf{Y}_\text{p} - \mathcal{U}(\mathbf{Y}_\text{p},\mathbf{M}) \right) \odot \mathbf{M}.
\label{eq:denoise}
\end{equation}
The input to the CE head $\mathcal{U}$ is formed by channel-wise concatenation of the real part, the imaginary part, and the binary mask as follows:
  $\mathbf{X}_0 = \mathrm{concat}(\Re(\mathbf{Y}_\text{p}), \Im(\mathbf{Y}_\text{p}), \mathbf{M}) \in \mathbb{R}^{3 \times M_y \times M_z}$.
Following an initial $3{\times}3$ Convolution-BatchNorm-ReLU (CBR) layer for feature extraction, the network adopts a U-Net architecture. Its basic building block consists of two consecutive $3{\times}3$ CBR operations followed by a dropout layer. The encoder sequentially applies four such blocks, each followed by $2{\times}2$ max-pooling, leading to a bottleneck block. Symmetrically, the decoder comprises four stages. At each stage, a $2{\times}2$ transposed convolution doubles the spatial resolution, the upsampled feature is concatenated with the corresponding encoder skip connection, and the result is processed by another building block. Finally, a $1{\times}1$ convolution maps the decoded features to the predicted noise residual, from which the denoised channel is recovered via~\eqref{eq:denoise}.

Subsequently, the denoised partial channel $\hat{\mathbf{H}}_\text{p}$ serves as the conditioning input to the foundation model, which extrapolates the full channel $\hat{\mathbf{H}}_0$. For fine-tuning, we adopt an end-to-end joint optimization strategy. The gradients are backpropagated from the foundation model through $\hat{\mathbf{H}}_\text{p}$ to update the parameters of the CE head, allowing the denoising head to learn features that are better aligned with the generative priors of the foundation model. To improve training stability, the total objective is defined as
\begin{equation}
    \mathcal{L}_\text{total} =
    \mathcal{L}_\text{diff}(\boldsymbol{\epsilon}, \hat{\boldsymbol{\epsilon}}_\theta)
    + 0.1\, \mathcal{L}_\text{aux}(\mathbf{H}_0, \hat{\mathbf{H}}_{\text{p}}),
\end{equation}
where $\mathcal{L}_\text{diff}$ is the noise-prediction MSE for the diffusion process, and $\mathcal{L}_\text{aux}$ enforces denoising accuracy on the observed pilots:
\begin{equation}
    \mathcal{L}_\text{aux}(\mathbf{H}_0, \hat{\mathbf{H}}_\text{p})
    =
    \left\| (\mathbf{H}_0 \odot \mathbf{M}) - \hat{\mathbf{H}}_\text{p} \right\|_2^2 .
\end{equation}
By minimizing this joint objective, the network simultaneously learns pilot denoising and full-channel extrapolation.

\subsection{Channel Exploitation Tasks}

For channel exploitation tasks, the objective is to extract high-level task-relevant information directly from the learned channel representations. We design lightweight heads to map the pretrained backbone features to task-specific outputs, considering two representative tasks: near-field and far-field UE classification and 3D user localization. 
To balance computational efficiency with task-specific adaptation, we adopt a partial fine-tuning strategy. Specifically, we freeze the first $B$ Transformer blocks of the foundation model to retain the transferable channel features learned during pretraining, and only update the last $L$ blocks. These trainable blocks are jointly optimized with the proposed time feature fusion module and the task-specific heads described below.

\subsubsection{Multi-time feature fusion}

\begin{figure}[!t]
    \centering
    \includegraphics[width=0.98\linewidth]{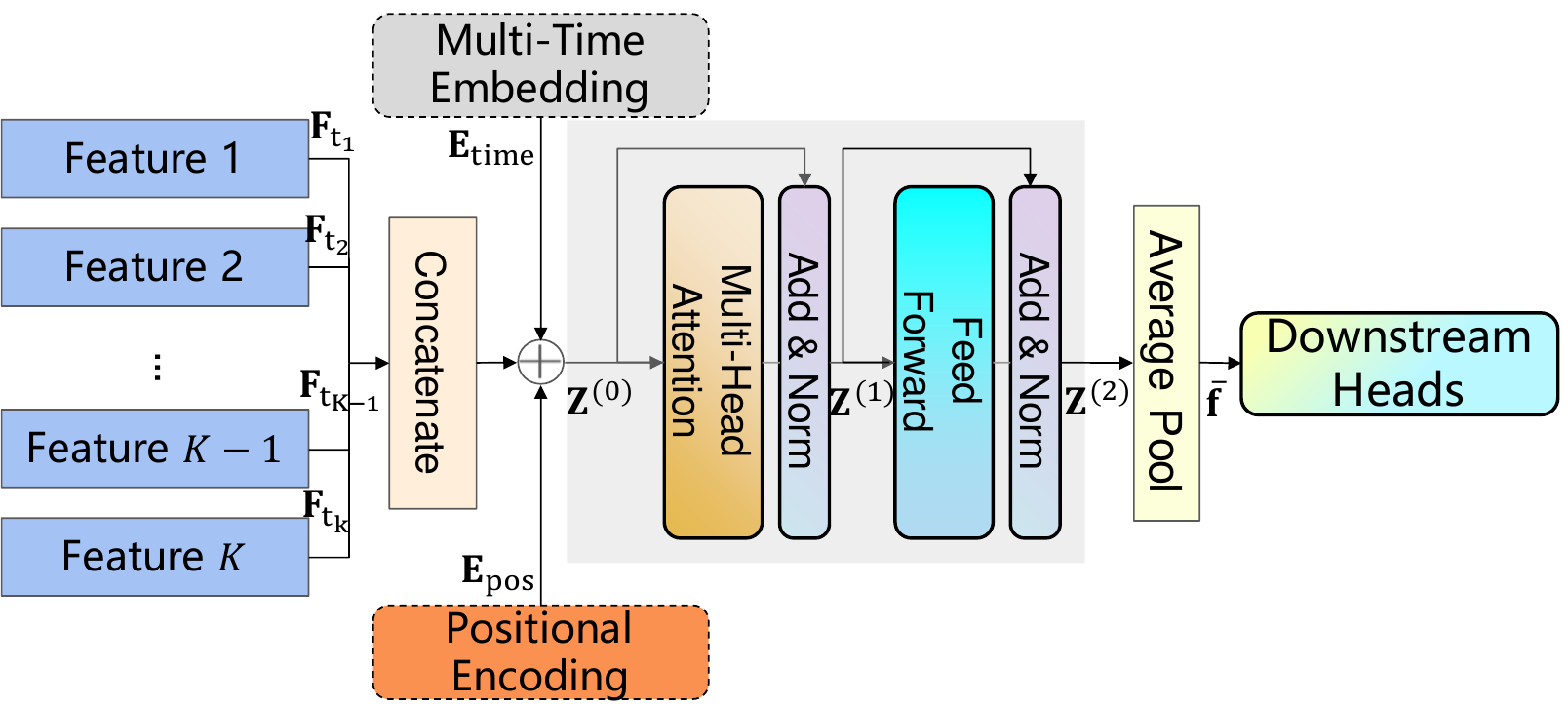}
    \caption{Attention-based multi-time feature fusion. Features from multiple diffusion time steps are fused through a MHA block over the concatenated token sequence, generating a global representation for downstream prediction.}
    \label{fig:fusion}
\end{figure}

As Fig.~\ref{fig:fusion} shows, the diffusion-based foundation model produces feature representations of different levels at different diffusion time steps~\cite{Luo_DiffusionHyperfeatures_2023, Tang_EmergentCorrespondence_2023}. In this work, we extract features from multiple selected diffusion time steps to capture complementary information across different noise levels. The corresponding feature maps are denoted by $\{\mathbf{F}_{t_j}\}_{j=1}^{T_s}$, where $T_s$ is the number of selected time steps and $\mathbf{F}_{t_j}\in\mathbb{R}^{N_{\text{tok}}\times D_f}$ contains $N_{\text{tok}}$ patch tokens of dimension $D_f$. To preserve both spatial and temporal contexts, these features are concatenated along the token dimension and augmented with learnable embeddings:
\begin{equation}
    \mathbf{Z}^{(0)} = \big[\mathbf{F}_{t_1},\,\mathbf{F}_{t_2},\,\dots,\,\mathbf{F}_{t_{T_s}}\big] + \mathbf{E}_{\text{pos}} + \mathbf{E}_{\text{time}} \in \mathbb{R}^{(T_s N_{\text{tok}})\times D_f},
\end{equation}
where $\mathbf{E}_{\text{pos}}\in \mathbb{R}^{(T_s N_{\text{tok}})\times D_f}$ and $\mathbf{E}_{\text{time}}\in \mathbb{R}^{(T_s N_{\text{tok}}) \times D_f}$ denote the spatial positional encoding and diffusion-step embedding, respectively. They allow the fusion module to distinguish both antenna-patch locations and diffusion time steps.

To model joint spatial and inter-time correlations, $\mathbf{Z}^{(0)}$ is processed by a standard Transformer encoder layer~\cite{vaswani2017attention}, utilizing a MHA block and a Feed-Forward Network (FFN) with pre-layer normalization (LN):
\begin{subequations}
\begin{align}
    \mathbf{Z}^{(1)} &= \mathbf{Z}^{(0)} + \operatorname{MHA}\big(\operatorname{LN}(\mathbf{Z}^{(0)})\big), \\
    \mathbf{Z}^{(2)} &= \mathbf{Z}^{(1)} + \operatorname{FFN}\big(\operatorname{LN}(\mathbf{Z}^{(1)})\big).
\end{align}
\end{subequations}
The MHA module captures dependencies across both spatial locations and diffusion time steps. Finally, we apply global average pooling over all $T_s N_{\text{tok}}$ tokens to condense the sequence into a unified global representation $\bar{\mathbf{f}} \in \mathbb{R}^{D_f}$:
\begin{equation}
    \bar{\mathbf{f}} = \frac{1}{T_s N_{\text{tok}}} \sum_{i=1}^{T_s N_{\text{tok}}} \mathbf{Z}^{(2)}_{i}.
\end{equation}
This fused vector $\bar{\mathbf{f}}$ serves as the input to task-specific heads, such as a classifier for near/far-field discrimination or a regressor for 3D coordinate estimation. The entire pipeline, comprising the unfrozen top backbone blocks, the fusion network, and the downstream heads, is fine-tuned end-to-end using supervised objectives on the downstream datasets.

\subsubsection{Task 3: User classification}

Near-/far-field user classification is formulated as a binary classification problem. The classification (CLS) head applies a normalization layer followed by a linear projection to map the fused feature $\bar{\mathbf f}$ to the logits
\begin{equation}
    \mathbf z^{(\mathrm{cls})}
    =
    \mathbf W_{\mathrm{cls}} \operatorname{Norm}(\bar{\mathbf f}) + \mathbf b_{\mathrm{cls}}
    \in \mathbb R^{2},
\end{equation}
where $\operatorname{Norm}(\cdot)$ denotes a normalization layer. The logits $z^{(\mathrm{cls})}_0$ and $z^{(\mathrm{cls})}_1$ correspond to the far-field and near-field classes, respectively, and are converted into class probabilities $\hat p^{(\mathrm{cls})}_{i,c}$ via the softmax function. For fine-tuning, we adopt the focal loss~\cite{lin2017focal}:
\begin{equation}
\mathcal L_{\mathrm{focal}}^{(\mathrm{cls})}
= \frac{1}{N_s}\sum_{i=1}^{N_s}\sum_{c=0}^{1}
\alpha_{c}\,(1-\hat p^{(\mathrm{cls})}_{i,c})^{\gamma_f}\,
\big(-\,y^{(\mathrm{cls})}_{i,c}\,\log \hat p^{(\mathrm{cls})}_{i,c}\big),
\end{equation}
where $y^{(\mathrm{cls})}_{i,c} \in \{0,1\}$ is the ground-truth (GT) label for the $i$-th sample, $\alpha_c$ is the class weight, and $\gamma_f$ is the focusing parameter.

\subsubsection{Task 4: User localization}

The user localization task is formulated as a regression problem for estimating the distance $r_{1,0}$ and the AoA $(\theta_{1,0}^{\mathrm R}, \phi_{1,0}^{\mathrm R})$ of the LoS path, as defined in Sec.~\ref{sec:system_model}. In this paper, we focus on a constrained LoS-dominant near-field localization setting to isolate the effect of the learned channel representation.
For simplicity, these quantities are denoted by $r$, $\theta$, and $\phi$, respectively.
The multilayer perceptron (MLP)-based localization (LOC) head outputs $(\hat{c}_1,\hat{s}_1,\hat{c}_2,\hat{s}_2,\hat{r}_{\text{raw}})$,
where $(\hat{c}_1,\hat{s}_1)$ and $(\hat{c}_2,\hat{s}_2)$ parameterize the two angles using cosine-sine pairs. The final predicted angles and distance are obtained through the following post-processing.

We first convert the cosine-sine pairs into raw angles:
\begin{subequations}
\begin{align}
  \hat{\theta}_{\text{raw}}
  &= 2\,\arctan\!\left(\frac{\hat{s}_1}{\sqrt{\hat{c}_1^{\,2}+\hat{s}_1^{\,2}}+\hat{c}_1}\right)\in(-\pi,\pi],\\
  \hat{\phi}_{\text{raw}}
  &= 2\,\arctan\!\left(\frac{\hat{s}_2}{\sqrt{\hat{c}_2^{\,2}+\hat{s}_2^{\,2}}+\hat{c}_2}\right)\in(-\pi,\pi].
\end{align}
\end{subequations}
The final predicted angles are then obtained by wrapping the raw angles into the target ranges:
\begin{subequations}
\begin{align}
  \hat{\phi}
  &\;=\; |\,((\hat{\phi}_{\text{raw}}+\pi)\bmod 2\pi)-\pi\,|
  \;\in[0,\pi], \\
  \hat{\theta}
  &\;=\; ((\hat{\theta}_{\text{raw}}+\tfrac{\pi}{2})\bmod \pi)-\tfrac{\pi}{2}
  \;\in[-\tfrac{\pi}{2},\,\tfrac{\pi}{2}].
\end{align}
\end{subequations}
For the distance prediction, the raw output $\hat r_{\text{raw}}$ is mapped into the target range through
\begin{equation}
    \hat r \;=\; r_{\min} +
    \frac{r_{\max}-r_{\min}}{1+e^{-\hat r_{\text{raw}}}}.
\end{equation}

We use separate distance and angle losses and combine them into a weighted localization loss. For $N_s$ samples with predicted distances $\{\hat r_i\}_{i=1}^{N_s}$ and GT distances $\{r_i\}_{i=1}^{N_s}$, the distance loss is defined as
\begin{equation}
\label{eq:dist-loss}
\mathcal{L}_{\text{dist}}
= \frac{1}{N_s}\sum_{i=1}^{N_s}\!\left(
\gamma_{\text{abs}}\, H_\delta\!\big(\hat r_i - r_i\big)
+\gamma_{\text{rel}}\, \frac{\lvert \hat r_i - r_i\rvert}{r_i}
\right),
\end{equation}
where $H_\delta(\cdot)$ denotes the Huber loss function~\cite{Huber_Robust_1964}, and $\gamma_{\text{abs}}$ and $\gamma_{\text{rel}}$ weight the absolute and relative distance errors, respectively.
For the angle loss, we match the cosine-sine representations of the predicted and GT angles. Given the predicted angles $\{\hat{\theta}_i,\hat{\phi}_i\}_{i=1}^{N_s}$ and the GT angles $\{\theta_i,\phi_i\}_{i=1}^{N_s}$, the loss is defined as
\begin{multline}
\label{eq:ang-vec}
\mathcal L_{\text{ang}}
= \frac{1}{N_s}\sum_{i=1}^{N_s}\!\big\|[\cos(\theta_i),\sin(\theta_i)]
- [\cos(\hat{\theta}_i),\sin(\hat{\theta}_i)]\big\|_2^2\\
\quad+\; \frac{\gamma_{\text{ele}}}{N_s}\sum_{i=1}^{N_s}\!\big\|[\cos(\phi_i),\sin(\phi_i)]
- [\cos(\hat{\phi}_i),\sin(\hat{\phi}_i)]\big\|_2^2,
\end{multline}
where $\gamma_{\text{ele}}$ controls the relative contribution of the elevation-angle term. The overall localization loss is
\begin{equation}
\mathcal L_{\text{loc}} = \mathcal L_{\text{dist}} + \gamma_{\text{ang}}\,\mathcal L_{\text{ang}},
\end{equation}
where $\gamma_{\text{ang}}$ balances the distance and angle losses.

\section{Experimental Results}
\label{sec:experimental results}

In this section, we evaluate the proposed framework on the \textbf{LAETwin-XL} dataset. We first describe the experimental settings and then analyze the experimental results.

\subsection{Experimental Setups}

\subsubsection{LAE Dataset Configuration}
The BS UPA consists of $M_z = 64$ elements in the vertical dimension and $M_y = 64$ elements in the horizontal dimension. The array center $\mathbf{o}_{\text{R}}$ is located at $(0,0,35\,\mathrm{m})$ in urban scenarios and at $(0,0,65\,\mathrm{m})$ in suburban scenarios. All antenna characteristics and channel parameters follow the 3GPP TR~38.901 specification. The carrier frequency is set to $f_c=7$~GHz, and the subcarrier spacing is 30~kHz. 
Since this work focuses on the 3D spatial characteristics of XL-MIMO channels rather than temporal evolution, the dataset is sampled at $\Delta\tau=1$~s. With $v_{\max}=20$~m/s and $f_c=7$~GHz, the maximum Doppler shift is $f_{\mathrm D,max}=v_{\max}f_c/c$, corresponding to a Doppler time scale of $1/f_{\mathrm D,max}\approx2.14$~ms. Since $\Delta\tau$ is much larger than this time scale, different snapshots can be regarded as approximately independent channel realizations. Within each snapshot, $T_{\text{sym}}\approx33.3~\mu$s is much smaller than the Doppler time scale, and thus the channel is treated as quasi-static over one OFDM symbol. Unless otherwise specified, we use the channel at the carrier frequency and the first OFDM symbol of each snapshot in the experiments. Moreover, the toolchain can also support temporal channel studies by using smaller snapshot intervals.

We generate two datasets: a large-scale pretraining dataset for foundation-model pretraining and a small downstream dataset for task-specific fine-tuning. A summary of both datasets is provided in Table~\ref{tab:datasets}. Each dataset is divided into training, validation, and test splits, and spans multiple geographic regions, including major urban centers (e.g., Beijing and Shanghai) as well as several suburban environments.
Within each region, we define multiple subscenes. For each subscene, we sample trajectories and channels for multiple UAVs with randomly initialized positions over $N_{\tau}=10$ consecutive time slots.
Table~\ref{tab:datasets} also gives the number of valid samples and the near-/far-field sample distribution in each subscene.
The total number of valid samples varies across cities due to the diverse topologies of the OSM-based real-world scenarios. For example, densely built environments such as urban New York generate fewer valid samples because severe physical blockages more frequently eliminate all propagation paths between the UE and the BS. In addition, some scenarios, such as urban London and Tokyo and suburban Singapore, exhibit an imbalanced near-/far-field distribution. This phenomenon is mainly related to the size of the sampled ROIs, in that most UE-to-BS distances remain within the Rayleigh distance in smaller ROIs, which leads to a disproportionately large number of near-field samples. Rather than indicating a limitation of the dataset, this imbalance reflects the spatial heterogeneity of real-world XL-MIMO propagation and provides challenging and realistic conditions for evaluating algorithmic robustness.

\begin{table}[!t]
\centering
\caption{Low-altitude XL-MIMO dataset splits, scene configurations, and sample statistics ($64{\times}64$ UPA).}
\label{tab:datasets}
\footnotesize
\renewcommand{\arraystretch}{0.9} 
\setlength{\tabcolsep}{3.5pt}  
\begin{tabular}{@{} l l c c c c @{}}
\toprule
\multirow{2}{*}{Split} &
\multirow{2}{*}{Scene (City)} &
\multirow{2}{*}{Scenes} &
\multicolumn{3}{c}{Hybrid-field channels} \\
\cmidrule(l){4-6}
& & & Total & Near & Far \\

\midrule
\multicolumn{6}{@{}l}{\textbf{Pre-training Large Dataset}} \\
\multirow{8}{*}{Train} 
& Urban (Shanghai)    & 10 & 9990 & 7021 & 2969 \\
& Urban (Tianjin)     &  9 & 8990 & 6158 & 2832 \\
& Urban (Hangzhou)    &  7 & 6910 & 5471 & 1439 \\
& Urban (Shenzhen)    &  5 & 5000 & 2388 & 2612 \\
& Urban (New York)     &  5 & 3604 & 2705 &  899 \\
& Urban (London)      &  5 & 4880 & 4345 &  535 \\
& Urban (Tokyo)       &  5 & 4950 & 4538 &  412 \\
& Suburban (Sydney)   &  5 & 4982 & 3514 & 1468 \\
\multirow{2}{*}{Val} 
& Urban (Singapore)   &  2 & 1930 &  948 &  982 \\
& Suburban (Singapore)&  2 & 2000 & 1746 &  254 \\
\multirow{2}{*}{Test} 
& Urban (Dubai)       &  2 & 2000 & 1775 &  225 \\
& Suburban (Dubai)    &  2 & 2000 & 1683 &  317 \\
\midrule
\multicolumn{6}{@{}l}{\textbf{Downstream Small Dataset}} \\
\multirow{2}{*}{Train} 
& Urban (Beijing)     & 10 & 9950 & 4576 & 5374 \\
& Suburban (Beijing)  &  1 &  500 &  450 &   50 \\
\multirow{2}{*}{Val} 
& Urban (Berlin)      &  2 & 1000 &  873 &  127 \\
& Suburban (Berlin)   &  1 &  500 &  443 &   57 \\
\multirow{2}{*}{Test} 
& Urban (Nanjing)     &  2 & 1000 &  895 &  105 \\
& Suburban (Nanjing)  &  1 &  500 &  386 &  114 \\
\bottomrule
\end{tabular}

\vspace{3pt}
\parbox{\linewidth}{\scriptsize
\textbf{Altitudes:} $[z_{\min}, z_{\max}]=[30,50]$\,m (urban), $[z_{\min}, z_{\max}]=[60,80]$\,m (suburban).
\textbf{Locations:} Urban splits are randomly cropped within city limits, suburban from rural outskirts.
}
\end{table}

\subsubsection{Baseline Description}

To comprehensively evaluate the proposed framework, we consider three groups of baselines.

\begin{itemize}
    \item \textbf{Foundation-model and generative baselines:}
    To validate the advantage of the proposed generative pretraining paradigm, we compare with several representative alternatives. First, to isolate the impact of the generative diffusion objective, we consider the  \textbf{large wireless model (LWM)}~\cite{alikhani2024largewirelessmodellwm}, which adopts the same Transformer backbone as our method but uses a deterministic masked channel modeling (MCM) pretraining scheme. Second, to compare diffusion-based generation with adversarial learning, we include a \textbf{conditional generative adversarial network (CGAN)}~\cite{CGAN}. Third, to assess the benefit of the proposed Transformer-based spatial modeling, we evaluate a \textbf{U-Net}~\cite{unet} variant in place of the proposed backbone.

    \item \textbf{Task-specific downstream baselines:}
    To demonstrate the effectiveness of the proposed framework on downstream tasks, we compare with representative specialized methods. For channel estimation, we adopt \textbf{ChannelNet}~\cite{soltani2019channelnet}, a widely used convolutional neural network (CNN)-based baseline. For user localization, we compare with the classical orthogonal matching pursuit (\textbf{OMP}) algorithm~\cite{OMP}. We also include \textbf{train from scratch}, where the proposed architecture with the downstream heads is trained entirely from scratch, in order to quantify the benefit of the pretrained channel representations.

    \item \textbf{Performance reference baselines:}
    To provide reference floors, we consider two simple baselines. For channel extrapolation, we use the \textbf{partial observation} itself, i.e., zero-filling all unobserved antennas. For channel estimation, we use the classical \textbf{least-squares (LS)} estimator.
\end{itemize}

\subsubsection{Implementation Details of the Proposed Algorithm}

The foundation model adopts a Transformer backbone with depth $N_{\mathrm{blk}}=10$, followed by a linear projection and an unpatchify operation. During pretraining, the model is trained for 2000 epochs with a batch size of 256 using AdamW and an initial learning rate of $2\times10^{-4}$. The diffusion process uses a linear noise schedule $\{\beta_t\}_{t=1}^{T}$ from $1\times10^{-4}$ to $2\times10^{-2}$ over $T=1000$ steps, and CDDIM reverse sampling is performed with interval $\Delta_t=20$, i.e., the number of sampling steps is 50. The mask ratio $\rho_M$ is uniformly sampled from $\{0.2,0.3,\ldots,0.8\}$, and the corresponding antenna selection ratio is $\rho_s=1-\rho_M$.
For downstream fine-tuning, all tasks are trained for 300 epochs using AdamW with weight decay $1\times10^{-4}$ unless otherwise specified. To enable parameter-efficient adaptation, we freeze the first $B=7$ Transformer blocks of the pretrained backbone and update only the last $L=3$ blocks. The unfrozen backbone blocks are optimized with learning rate $2\times10^{-5}$, while the downstream task-specific modules use learning rate $2\times10^{-4}$. 

The evaluation and downstream adaptation task settings are summarized as follows. For \textbf{channel extrapolation}, the pretrained model directly acts as an extrapolator to recover missing UPA entries under random masks. For \textbf{channel estimation}, performance is evaluated under different mask ratios and SNR conditions. For \textbf{user classification and user localization} tasks, we extract $N_{\text{tok}}=64$ patch tokens, and each of dimension $D_f=256$. In the proposed multi-timestep feature fusion, we extract and aggregate representations from $T_s=5$ distinct diffusion timesteps, i.e., $t_1=0$, $t_2=200$, $t_3=400$, $t_4=600$, and $t_5=1000$. The classification task uses a single linear head, whereas the localization task uses an MLP-based regression head. The focal loss uses focusing parameter $\gamma_f=1.8$ with inverse-frequency class weights. For localization, the loss parameters are set to $\delta=30$, $\gamma_{\text{abs}}=0.8$, $\gamma_{\text{rel}}=0.2$, and $\gamma_{\text{ele}}=6.0$, while the angular weight $\gamma_{\text{ang}}$ decays from 1.0 to 0.5 after 200 epochs.

To ensure a fair comparison, all methods use the same task-specific dataset splits. Channel extrapolation is evaluated on the test split of the large pretraining dataset, which contains 4,000 samples. All downstream tasks are evaluated on the standard split of the downstream small dataset, whose test split contains 1,500 samples. The only exception is user localization, for which training and evaluation are conducted on the \emph{LoS-only near-field suburban subset} of the downstream small dataset, containing 450 training samples, 443 validation samples, and 386 test samples.
For pretraining-based baselines, including LWM, CGAN, and CDDIM w. U-Net, all models are pretrained from scratch on the same large pretraining dataset using the same training/validation split and pretraining protocol as the proposed method, i.e., 2,000 epochs, a batch size of 256, AdamW, and an initial learning rate of $2\times10^{-4}$, unless otherwise specified. For downstream tasks other than channel extrapolation, all learning-based methods are adapted under the same optimization settings to avoid discrepancies caused by hyperparameter tuning. Unless otherwise specified, the downstream fine-tuning epoch is 300. For ChannelNet, however, we retain its original two-stage training strategy, using 1,000 epochs for the super-resolution stage and 1,000 epochs for the restoration stage.
For the OMP baseline~\cite{Li_KeypointNF_Localization_2025}, we use a polar-domain codebook of size $D=729{,}000$, constructed from a $90\times90\times90$ grid uniformly sampled over azimuth $\theta\in[-\pi/2,\pi/2]$, elevation $\phi\in[0,\pi]$, and distance $d\in[10,300]$~m. All experiments are conducted on a workstation equipped with two NVIDIA GeForce RTX 4090 GPUs.

\subsection{Tasks Performance Evaluation}

We next present performance evaluations of the proposed method on the four tasks described above and compare them with baselines.

\subsubsection{Task 1: Channel Extrapolation}

\begin{figure}[!t]
    \centering
    \includegraphics[width=0.95\linewidth]{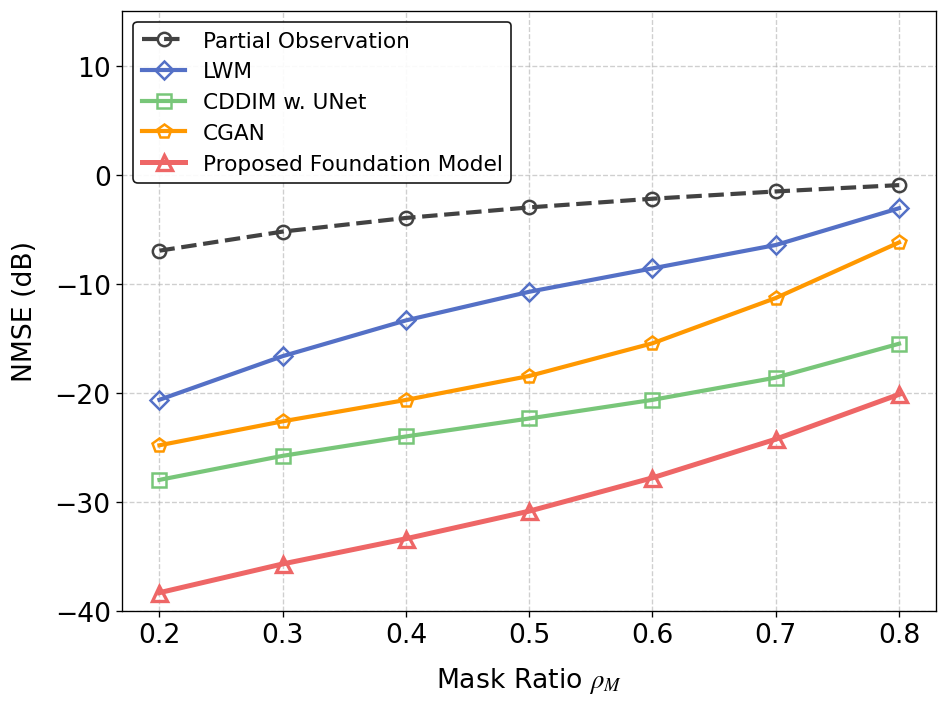}
    \caption{Comparison of channel extrapolation performance using different models under various mask ratios.}
\label{fig:results_t1}
\end{figure}

\begin{figure*}[!t]
    \centering
    \includegraphics[width=0.85\linewidth]{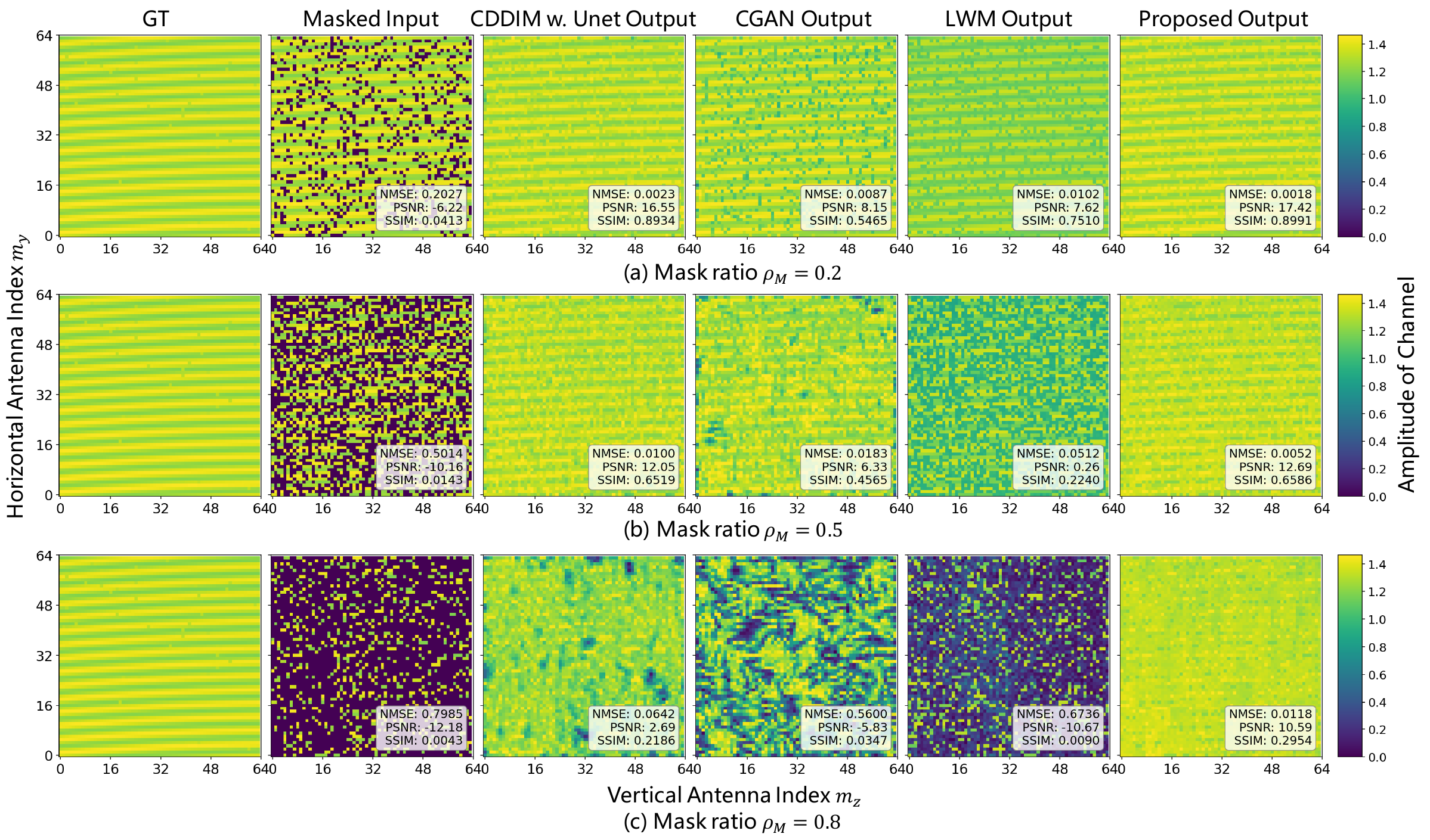}
    \caption{Visual comparison of extrapolated channel amplitudes generated by different models under various mask ratios.}
\label{fig:channel_amplitude}
\end{figure*}

As illustrated in Fig.~\ref{fig:results_t1}, the proposed foundation model consistently achieves the lowest normalized mean squared error (NMSE) across all mask ratios $\rho_M$. The performance gain can be attributed to three aspects of the proposed design. First, it substantially outperforms the MCM pretraining approach used in LWM by more than $15$~dB. This demonstrates that the generative pretraining paradigm is much more effective than MCM in the considered setting in capturing underlying distributions for robust channel extrapolation. Second, the performance gap over CGAN validates the efficacy and stability of the CDDIM algorithm in generating high-fidelity channels compared with adversarial training. Finally, our method surpasses the CDDIM with U-Net baseline, validating the advantage of the proposed physics-aware positional encoding combined with a Transformer backbone, which extracts and preserves complex channel features much more effectively than standard convolutional architectures. Notably, even with a high mask ratio of $0.8$, our model maintains an NMSE of approximately $-20$~dB. 
Moreover, Fig.~\ref{fig:channel_amplitude} visualizes the amplitudes of extrapolated channels under different mask ratios, providing a direct comparison of reconstruction fidelity and structural preservation. It can be observed that the proposed method is consistently visually the closest to the GT, preserving both the global spatial structure and fine-grained local details while avoiding the blurring and artifacts observed in the baselines. This visual advantage is consistent with the quantitative results, where the proposed method achieves the best NMSE, peak SNR (PSNR), and structural similarity index measure (SSIM).

\begin{figure*}[!t]
    \centering
    \includegraphics[width=0.8\linewidth]{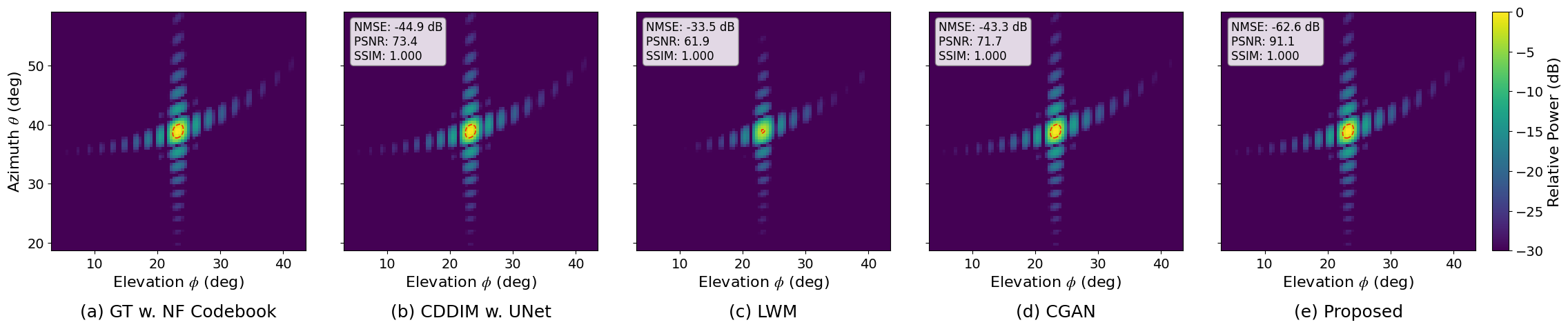}
    \caption{Angular-domain channel visualizations using different models ($\rho_M=0.5$). The red dashed lines are the -3\,dB contours.}
\label{fig:angular-domain}
\end{figure*}

\begin{figure}[!t]
    \centering
    \includegraphics[width=0.9\linewidth]{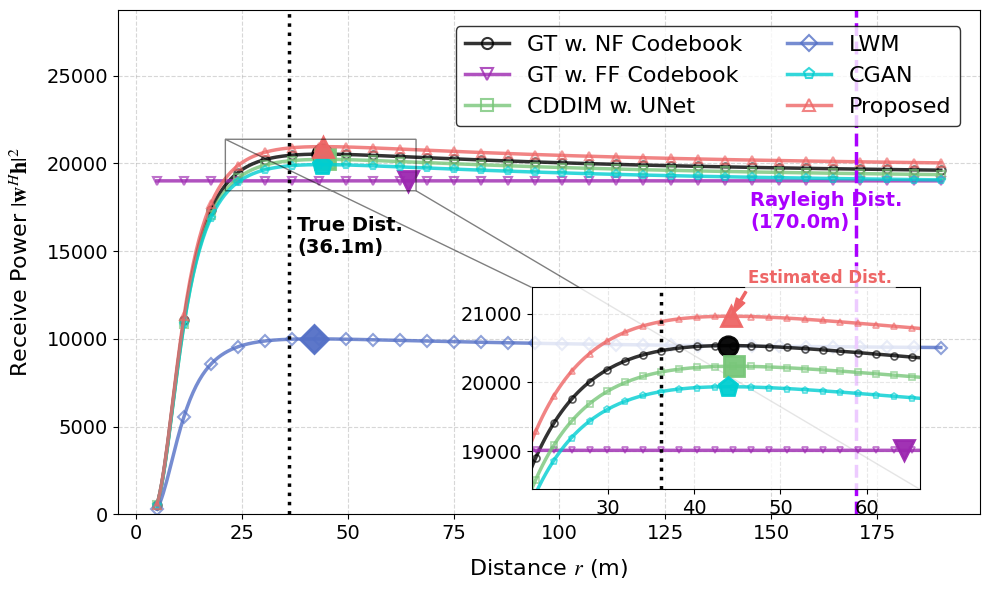}
    \caption{Polar-domain receive power distributions across distance dimension for extrapolated channels ($\rho_M=0.5$). The GT evaluated with a far-field codebook is also included to highlight the near-field distance resolution. Note: Markers are downsampled in the main plot for clarity, full density used in the codebook is shown in the inset.}
\label{fig:receive_power}
\end{figure}

\newcommand{\imgwidth}{0.32\linewidth}
\begin{figure*}[t] 
\centering

\subfigure[Antenna selection ratio $\rho_\text{s}=1.0$.]{%
    \includegraphics[width=\imgwidth]{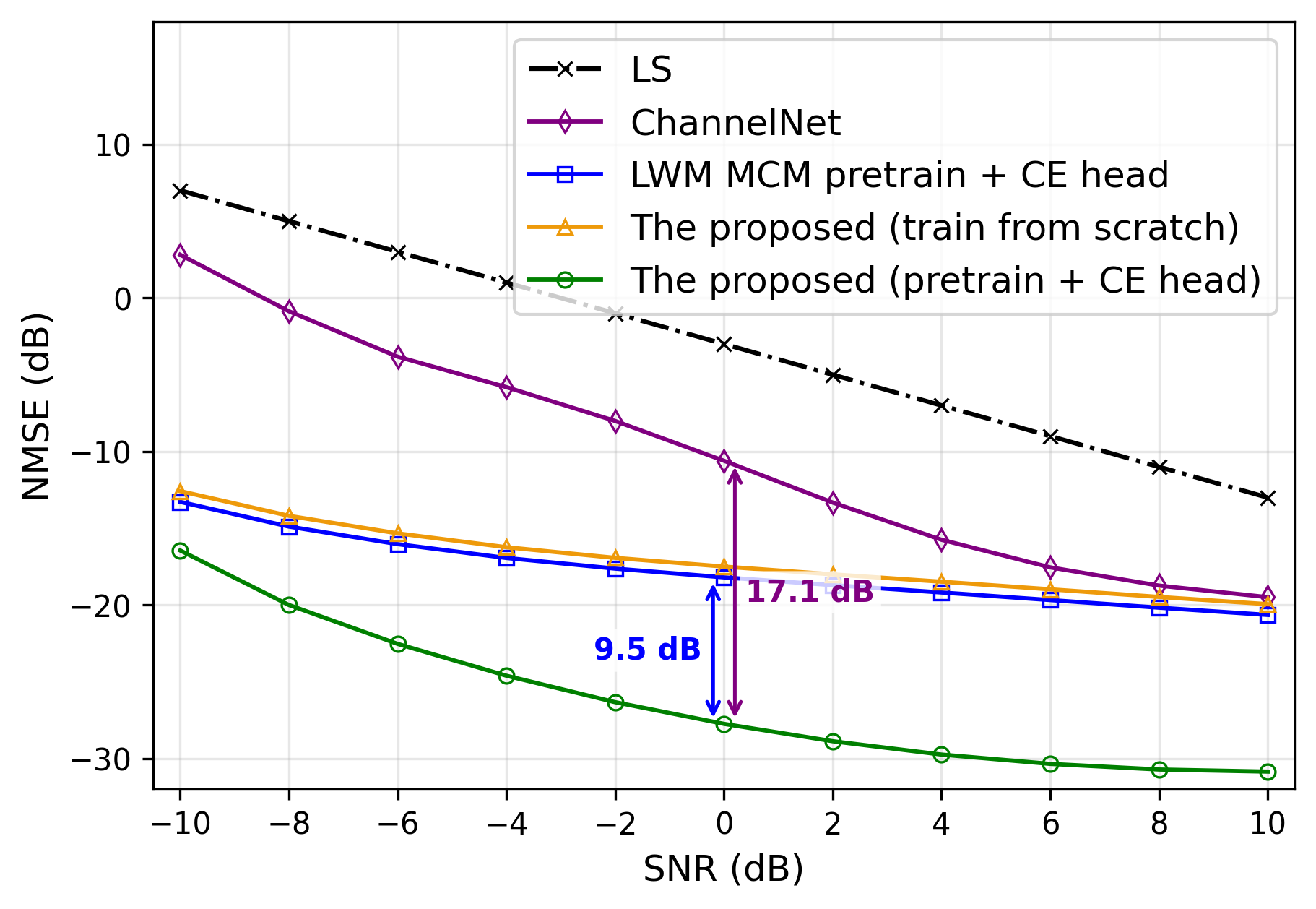}%
}\hspace{0\linewidth} 
\subfigure[Antenna selection ratio $\rho_\text{s}=0.5$.]{%
    \includegraphics[width=\imgwidth]{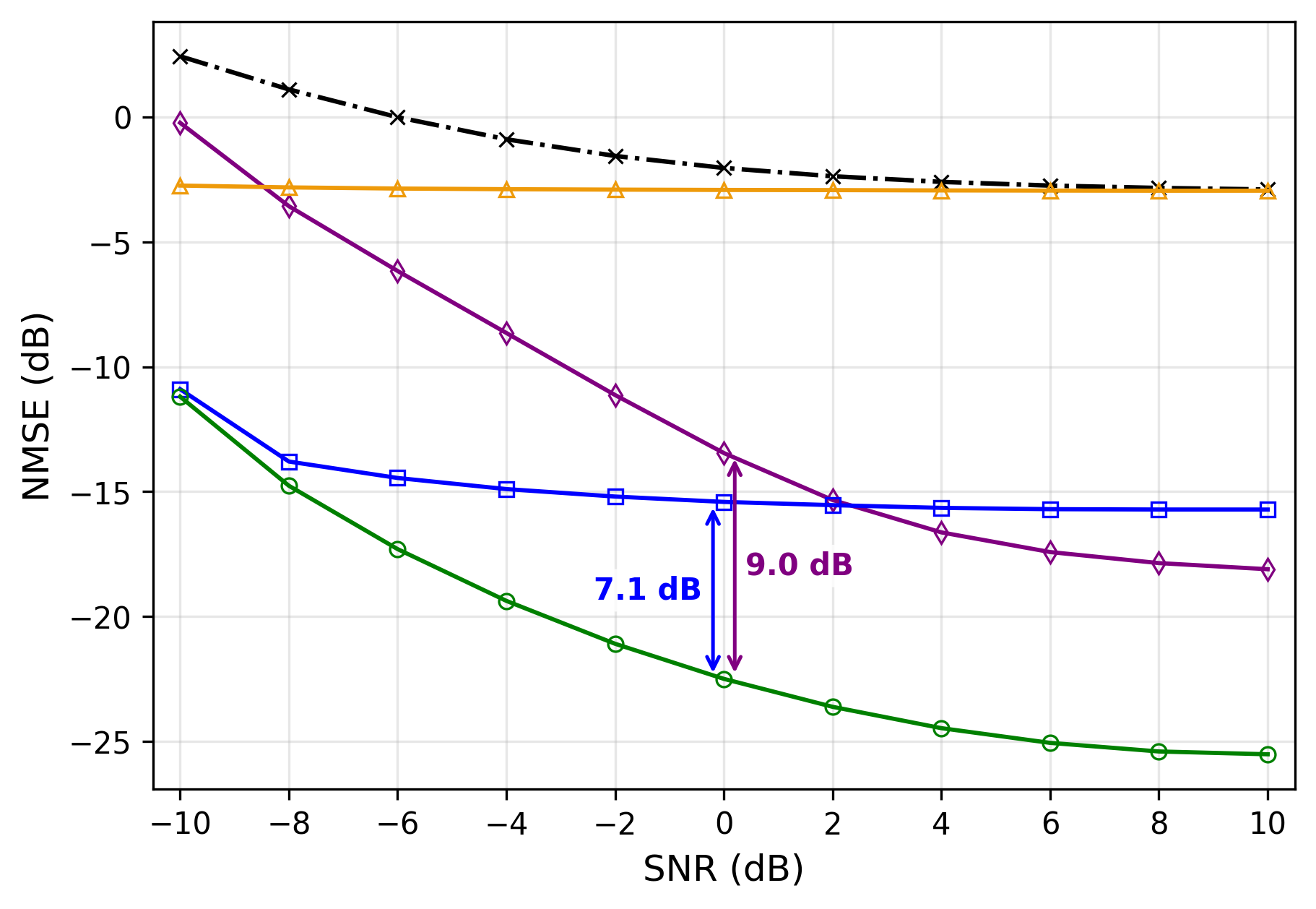}%
}\hspace{0\linewidth} 
\subfigure[Antenna selection ratio $\rho_\text{s}=0.2$.]{%
    \includegraphics[width=\imgwidth]{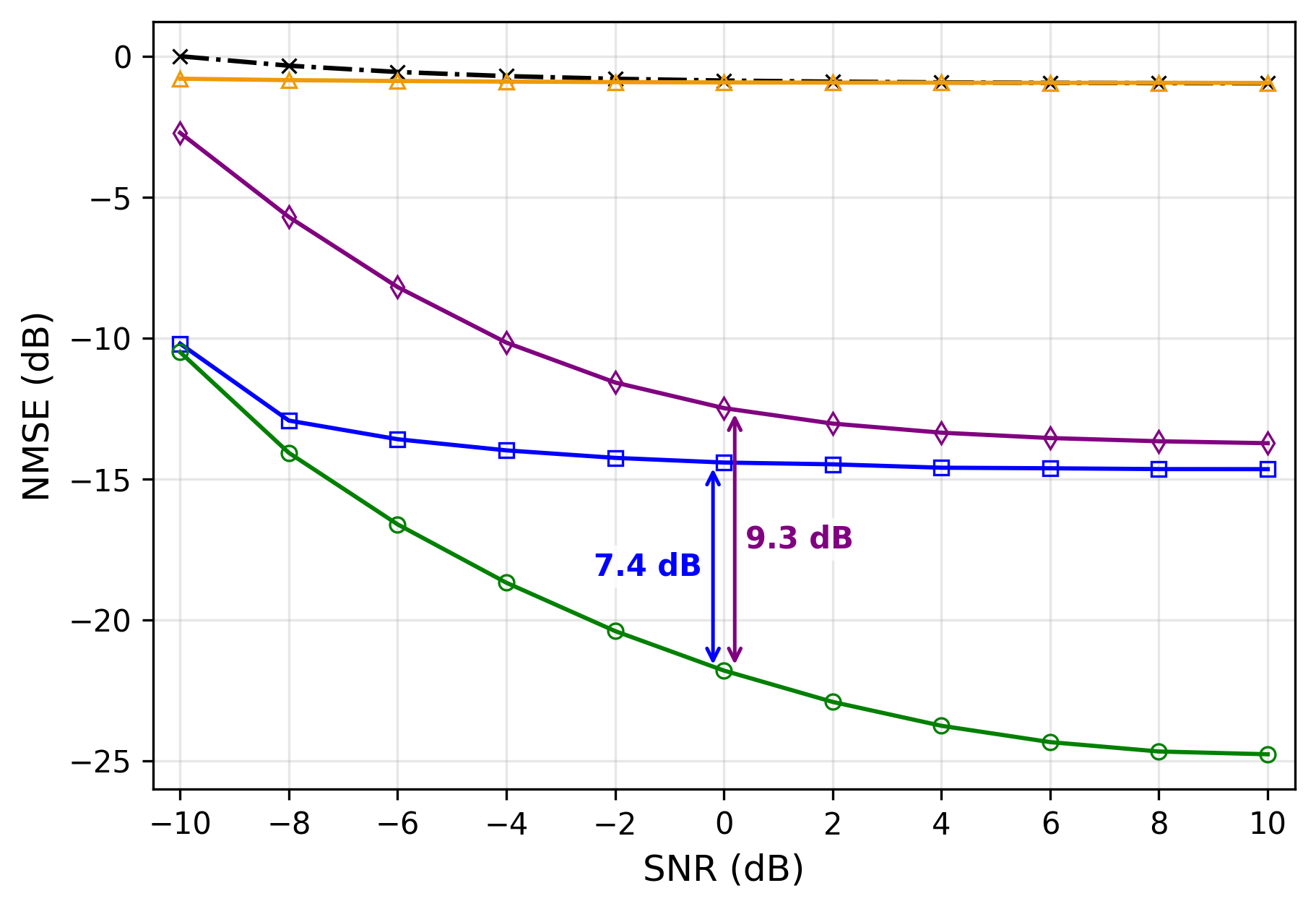}%
}

\caption{Comparison of channel estimation performance versus SNRs under different antenna selection ratios.}
\label{fig:results_ce_1}
\end{figure*}

\renewcommand{\imgwidth}{0.32\linewidth} 

\begin{figure*}[t]
\centering

\subfigure[SNR $= -10$ dB.]{%
    \includegraphics[width=\imgwidth]{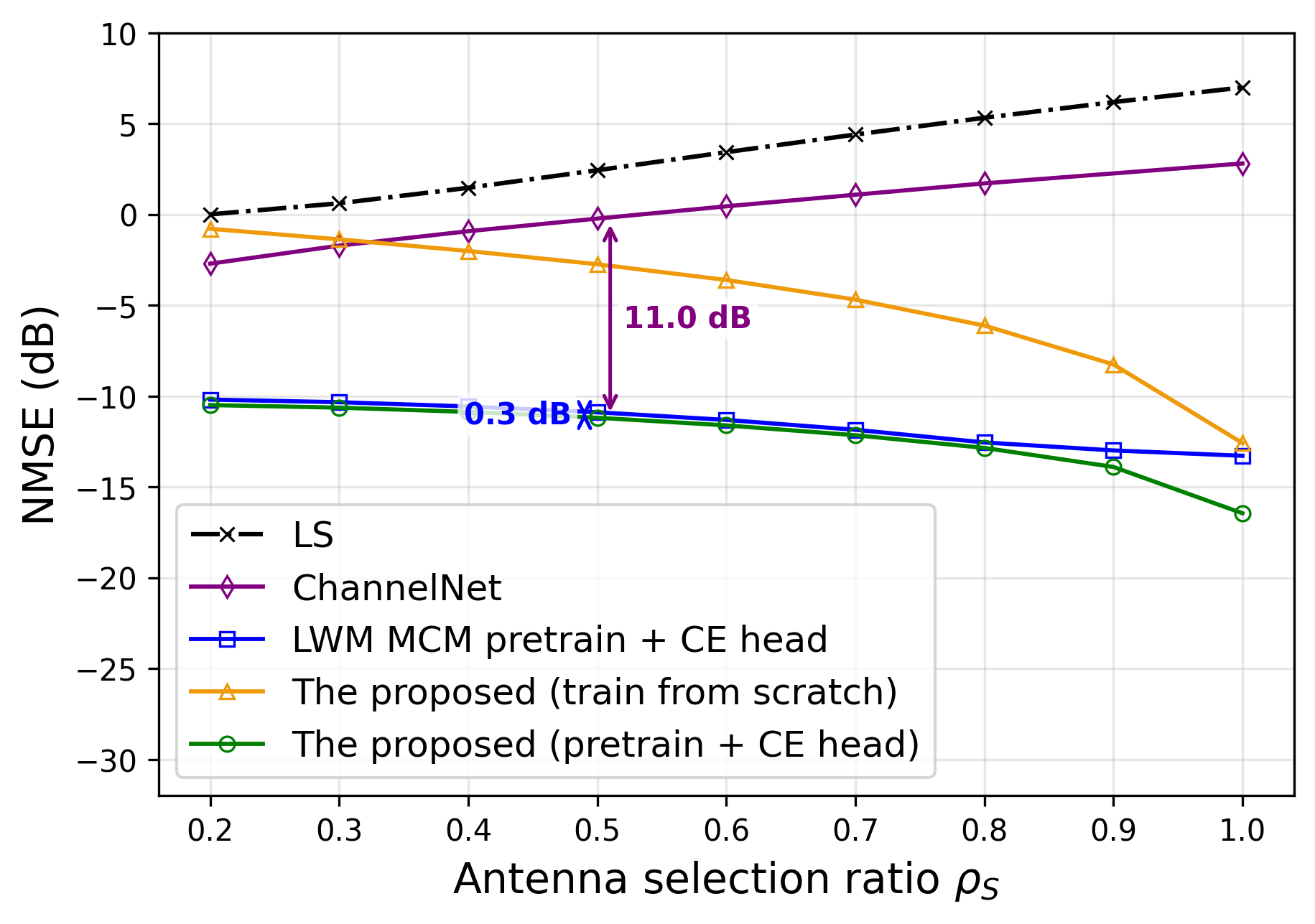}%
}\hspace{0\linewidth} 
\subfigure[SNR $= 0$ dB.]{%
    \includegraphics[width=\imgwidth]{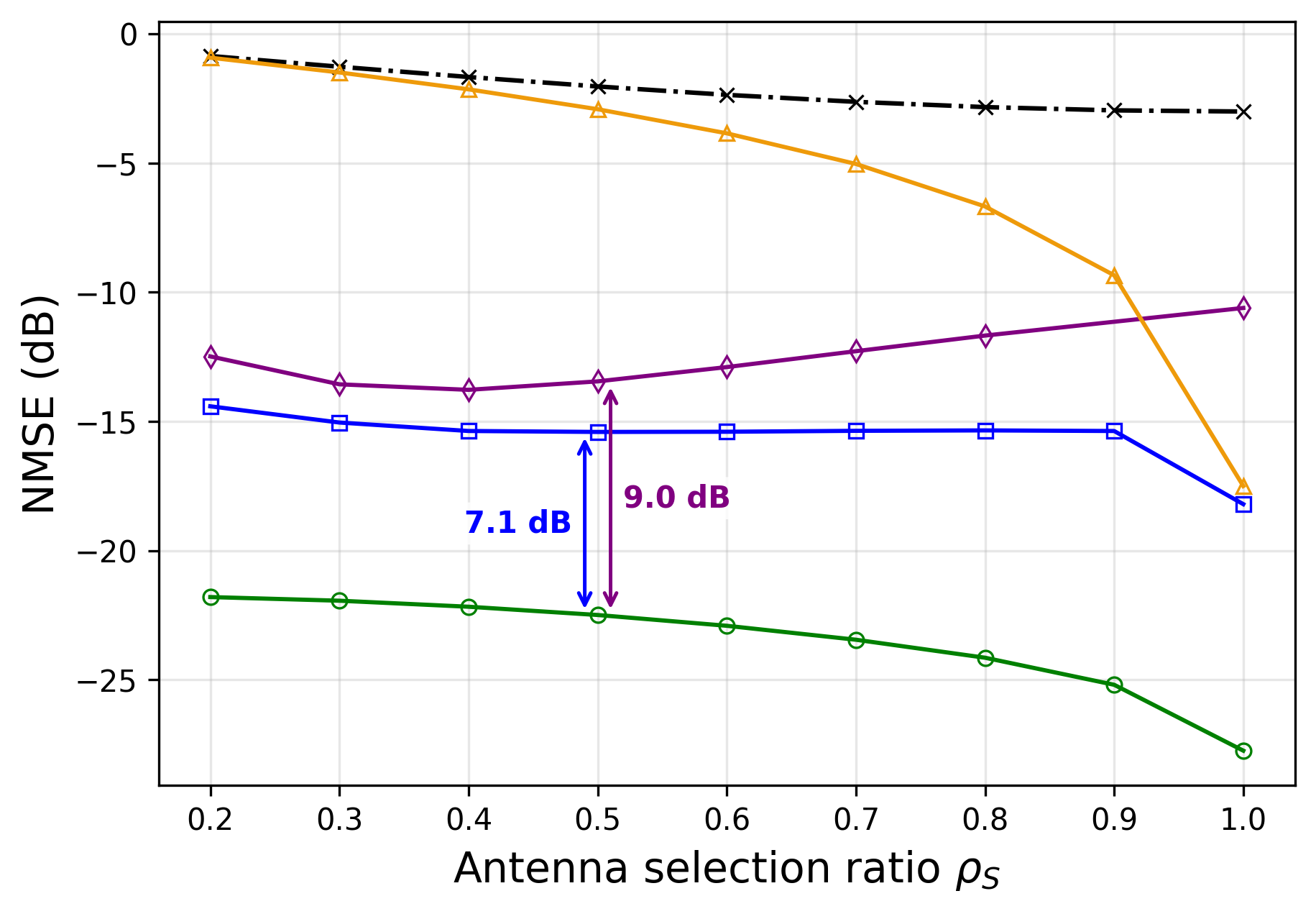}%
}\hspace{0\linewidth} 
\subfigure[SNR $= 10$ dB.]{%
    \includegraphics[width=\imgwidth]{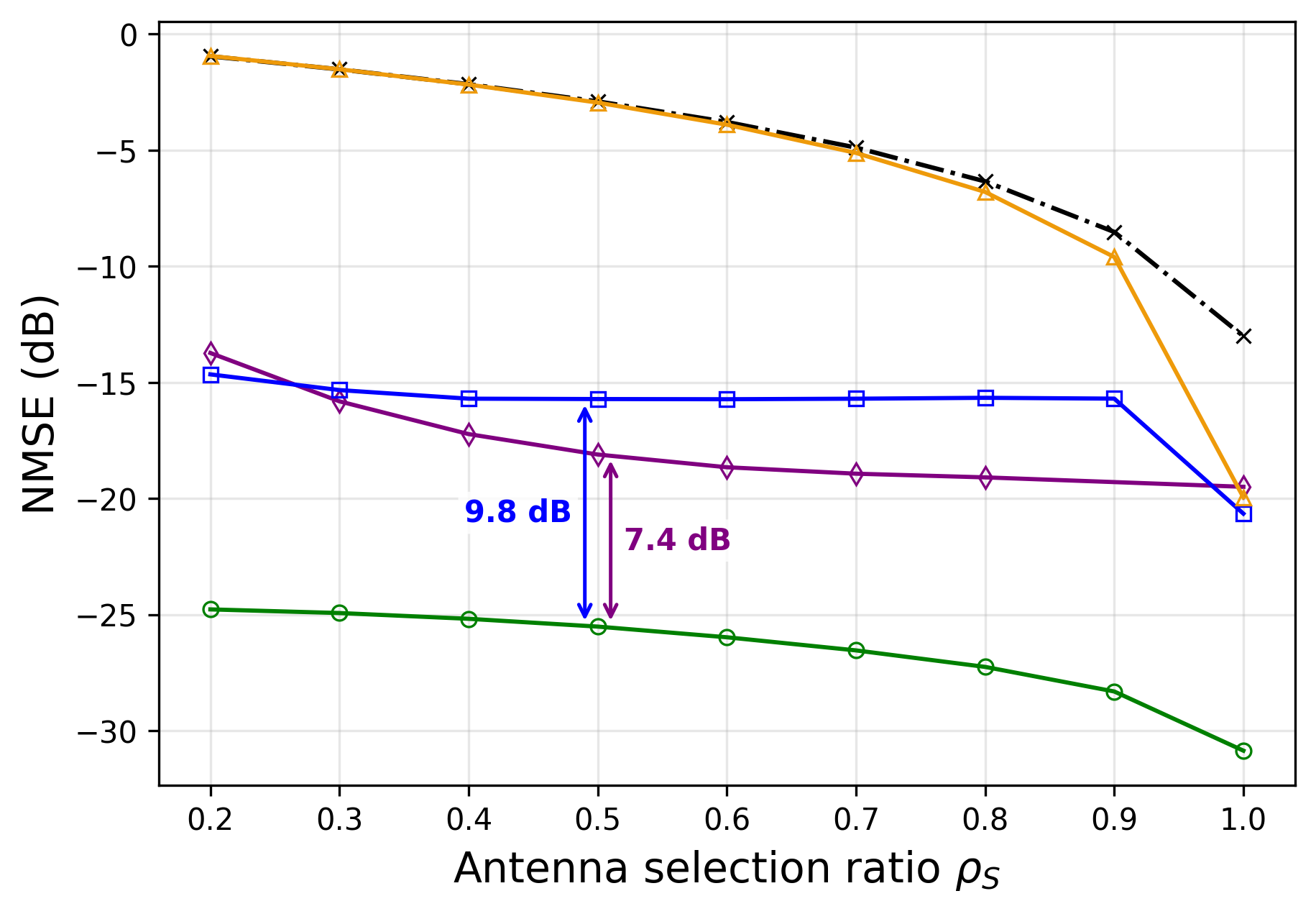}%
}

\caption{Comparison of channel estimation performance versus antenna selection ratio under different SNRs.}
\label{fig:results_ce_2}
\end{figure*}

We further examine whether the extrapolated channels remain physically consistent in the polar domain using the same polar-domain codebook as the OMP baseline. We consider a representative near-field user located 36.1~m from the BS as an illustrative example.
First, Fig.~\ref{fig:angular-domain} presents the polar-domain channel visualizations along the angular dimensions (elevation and azimuth) at $\rho_M = 0.5$. Among the baselines, LWM shows the weakest angular concentration, with severe energy dispersion and the $-3$~dB contour that deviates most from the GT. In contrast, the proposed method achieves the best quantitative performance in terms of NMSE, PSNR, and SSIM, while also exhibiting the closest main-lobe profile to the GT. This result indicates that the proposed model better preserves the angular sparsity of the underlying propagation paths.
Next, Fig.~\ref{fig:receive_power} shows the polar-domain receive power of the extrapolated channels over the distance dimension. Almost all models exhibit clear distance focusing, with the estimated distances (dominant peaks) appearing near the true user distance, indicating that the extrapolated channels retain the distance selectivity characteristic of near-field propagation. By contrast, when the GT channel is evaluated using a conventional far-field codebook, the receive power becomes nearly flat over the distance dimension, revealing the spatial energy spread caused by the channel model mismatch.
These results suggest that the proposed foundation model can capture key near-field propagation characteristics from partial observations, enabling the extrapolated full channels to better reconstruct spherical-wave characteristics.

\subsubsection{Task 2: Channel Estimation}

Fig.~\ref{fig:results_ce_1} first shows the NMSE versus SNR under three representative antenna selection ratios, i.e., $\rho_s = 1.0, 0.5,$ and $0.2$. Compared with the same architecture trained from scratch, the pretrained foundation model remains markedly more robust under severe channel information loss at low selection ratios (e.g., $\rho_s=0.5$ and $0.2$), which validates that the proposed generative pretraining provides a strong channel prior and a favorable initialization, which helps the model adapt effectively to downstream channel estimation under partial antenna observations. Furthermore, the proposed approach exhibits substantial advantages over the LWM MCM pretrained model with the same CE head, which validates the effectiveness of the proposed generative pretraining scheme. In addition, it significantly outperforms the CNN-based ChannelNet, as the utilized Transformer backbone can capture the global spatial correlations across the massive XL-MIMO array that local convolution kernels fail to observe.

Then, Fig.~\ref{fig:results_ce_2} investigates the NMSE performance versus the antenna selection ratio under SNR=$-10$, $0$, and $10$~dB, respectively. The proposed algorithm still maintains significant superiority across all scenarios. Compared with LWM, the performance gains are particularly pronounced at moderate to high SNRs ($0, 10$~dB). At $-10$~dB, the performance margin narrows slightly, as the estimation accuracy in this severe regime is predominantly dictated by the inherent capability of the denoising head rather than the pretrained prior. Moreover, it can be observed that at lower SNRs ($-10$ and $0$~dB), the NMSE of the LS and ChannelNet baselines does not always improve as the antenna selection ratio increases. This is mainly because these baselines have limited denoising capability. Under severe noise, increasing the antenna selection ratio also increases the number of noisy observed entries directly involved in the reconstructed channel. Without an effective generative prior, the additional observations may not translate into improved NMSE. In contrast, the proposed model benefits from both the CE head and the pretrained generative prior, enabling more stable performance improvement as the antenna selection ratio increases.

\begin{table*}[t]
\centering
\caption{Comparison of user classification accuracy (\%) under different mask ratios.}
\label{tab:user_cls}
\setlength{\tabcolsep}{9pt} 
\renewcommand{\arraystretch}{1.15}  
\begin{tabular}{lccccccccc}
\toprule
Method & 0.0 & 0.1 & 0.2 & 0.3 & 0.4 & 0.5 & 0.6 & 0.7 & 0.8 \\
\midrule
The proposed pre-train + CLS head
& \textbf{95.87} & \textbf{95.67} & \textbf{95.73} & \textbf{95.67} & \textbf{95.53} & \textbf{95.40} & \textbf{95.40} & \textbf{95.13} & \textbf{95.13} \\
LWM MCM pre-train + CLS head
& 90.73 & 90.40 & 90.40 & 90.13 & 90.13 & 90.47 & 90.47 & 90.18 & 89.40 \\
Train from scratch
& 85.93 & 85.53 & 85.00 & 84.67 & 84.33 & 83.27 & 81.47 & 81.33 & 79.20 \\
\bottomrule
\end{tabular}
\end{table*}

\subsubsection{Task 3: User Classification}

Table~\ref{tab:user_cls} evaluates the downstream performance on near-field and far-field user classification across varying mask ratios. With full channel observations ($\rho_M=0$), the proposed generative framework achieves the highest accuracy of $95.87\%$, outperforming the training-from-scratch and LWM MCM baselines by $9.94\%$ and $5.14\%$, respectively. More importantly, our scheme exhibits exceptional resilience to severe data incompleteness. Even at an extreme mask ratio of $\rho_M=0.8$, it sustains a remarkable accuracy of $95.13\%$, whereas the training-from-scratch baseline degrades drastically to $79.20\%$. This negligible performance drop (merely $0.74\%$) of the proposed method indicates that it can still capture informative channel features even when most antennas are masked.

\subsubsection{Task 4: User Localization}

\begin{figure*}[!t]
    \centering
    \includegraphics[width=0.99\linewidth]{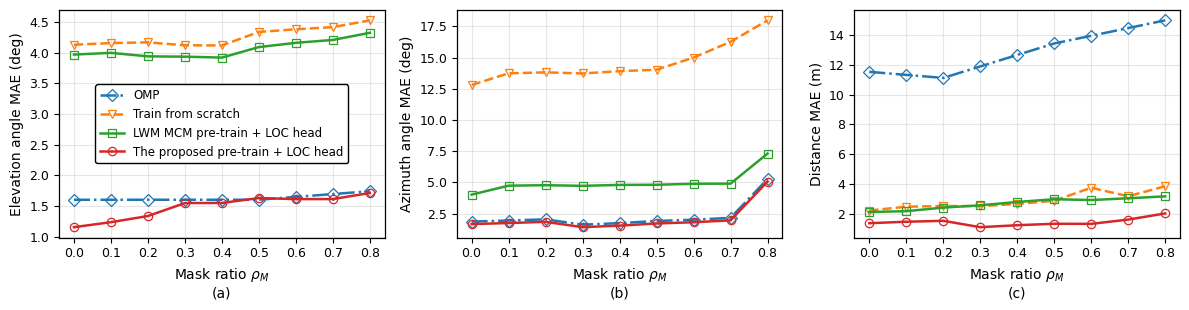}
    \caption{Comparison of localization performance under varying mask ratios for different algorithms, in terms of (a) elevation angle MAE, (b) azimuth angle MAE, and (c) distance MAE.}
    \label{fig:results_t5_1}
\end{figure*} 

Fig.~\ref{fig:results_t5_1} quantifies the mean absolute error (MAE) for elevation, azimuth, and distance estimation across varying mask ratios $\rho_M$. The proposed foundation model with LOC head consistently achieves the lowest and most stable MAE. Even at $\rho_M=0.8$, the elevation and azimuth MAEs remain within $2^{\circ}$ and $5^{\circ}$, respectively, while the distance error stays within $2$\,m. In contrast, although OMP attains comparable angular accuracy, it suffers from severe ranging errors ($11$--$17$\,m), mainly because its fixed discretized near-field codebook cannot precisely adapt to the actual propagation environment, and the off-grid effect further degrades distance estimation. The train-from-scratch baseline also exhibits large azimuth deviations (exceeding $12.5^{\circ}$), while LWM MCM remains consistently inferior to the proposed generative scheme.

\begin{figure}[!t]
    \centering
    \includegraphics[width=0.9\linewidth]{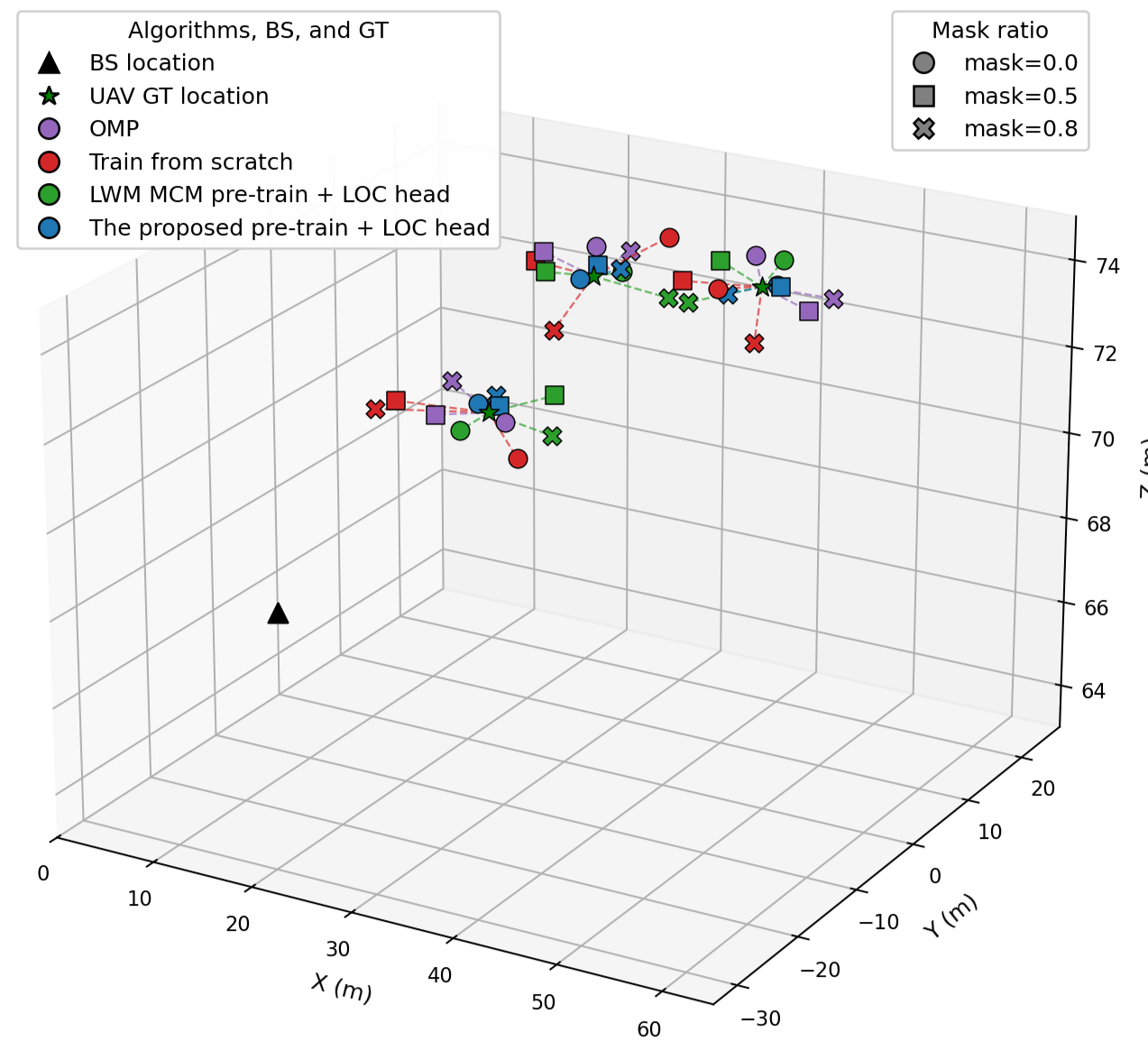}
    \caption{Visualization of UAV localization results.}
    \label{fig:results_t5}
\end{figure}

Fig.~\ref{fig:results_t5} further visualizes the 3D localization results for three representative UAV locations. At low mask ratios, all methods achieve reasonable accuracy since the partial observations still retain sufficient spatial information. As $\rho_M$ increases to $0.8$, however, the baselines degrade significantly, whereas the proposed method continues to produce estimates close to the GT. These results suggest that the proposed pretrained generative foundation model can capture a robust prior of near-field spatial geometry, thereby improving localization reliability under severe observation loss.

\begin{table}[t]
\centering
\footnotesize 
\renewcommand{\arraystretch}{0.9} 
\setlength{\tabcolsep}{0.8pt}
\caption{Ablation studies on key hyperparameters of the proposed algorithm.}
\begin{tabular}{lllc}
\toprule
\textbf{Category} & \textbf{Hyper-parameters} & \textbf{Information} & \textbf{Performance} \\
\midrule
\textbf{Diffusion}$^{\dag}$ 
 & $\Delta_t=5$ & 200 steps & -31.16 dB (289.91ms) \\
 & $\Delta_t=10$ & 100 steps & -30.83 dB (142.94ms) \\
\rowcolor{gray!20} & $\Delta_t=20$ & 50 steps & \textbf{-30.08 dB (51.02ms)}  \\
 & $\Delta_t=50$ & 20 steps & -27.25 dB (31.26ms) \\
 & $\Delta_t=100$ & 10 steps & -23.92 dB (19.59ms) \\
\midrule
\textbf{Depth}$^{\dag}$ 
 & $N_{\mathrm{blk}}=3$ & Size: 14.78 M & -24.75 dB \\
 & $N_{\mathrm{blk}}=7$ & Size: 31.59 M & -27.47 dB \\
\rowcolor{gray!20} & $N_{\mathrm{blk}}=10$ & Size: 44.19 M & \textbf{-30.08 dB} \\
 & $N_{\mathrm{blk}}=12$ & Size: 52.60 M & -30.91 dB \\
\midrule
\textbf{Fine-tuning}$^{\ddag}$ 
 & $(B,L)=(1,9)$ & -- & 89.89\% \\
 & $(B,L)=(5,5)$ & -- & 95.38\% \\
\rowcolor{gray!20} & $(B,L)=(7,3)$ & -- & \textbf{95.50\%} \\
 & $(B,L)=(9,1)$ & -- & 94.50\% \\
\midrule
\textbf{Fusion}$^{\ddag}$ 
 & $\{0\}$ & $T_s=1$ & 94.72\% \\
 & $\{200\}$ & $T_s=1$ & 94.51\% \\
 & $\{400\}$ & $T_s=1$ & 92.91\% \\
 & $\{600\}$ & $T_s=1$ & 92.28\% \\
 & $\{1000\}$ & $T_s=1$ & 94.43\% \\
 & $\{0,400,1000\}$ & $T_s=3$ & 94.85\% \\
\rowcolor{gray!20} & $\{0,200,400,600,1000\}$ & $T_s=5$ & \textbf{95.50\%} \\
\bottomrule
\end{tabular}

\vspace{4pt}
\parbox{\linewidth}{\scriptsize
$^{\dag}$ Evaluated on the channel extrapolation task under various mask ratios (measuring average NMSE and inference latency).\\
$^{\ddag}$ Evaluated on the downstream user classification task (measuring average Accuracy).
}
\label{tab:ablation_all}
\end{table}

\begin{table*}[!t] 
\centering
\caption{Component-wise and task-specific overhead analysis.}
\label{tab:overhead_comparison}
\footnotesize
\renewcommand{\arraystretch}{0.95}
\begin{tabular*}{\linewidth}{@{\extracolsep{\fill}} lccccc @{}}
\toprule
\multicolumn{6}{@{}l}{\textbf{Complexity of Each Component}} \\
\midrule
\textbf{Module:} & \textbf{Foundation Model} & \textbf{Multi-time Fusion} & \textbf{CE Head} & \textbf{CLS Head} & \textbf{LOC Head} \\
\textbf{Role:} & \textit{Shared backbone} & \textit{Shared extractor} & \multicolumn{3}{c}{\textit{Task-specific lightweight head}} \\
\textbf{Params / Size:} & 44.19 M / 168.6 MB & 10.25 M / 39.1 MB & \textbf{3.13 M} / \textbf{12.1 MB} & \textbf{0.003 M} / \textbf{0.1 MB} & \textbf{1.06 M / 4.2 MB} \\
\bottomrule
\end{tabular*}

\vspace{0.8em} 
\setlength{\tabcolsep}{3pt} 
\begin{tabular*}{\linewidth}{@{\extracolsep{\fill}} l l c c c c l @{}} 
\toprule
\textbf{Task} & \textbf{Method} & \textbf{Params (M)} & \textbf{Size (MB)} & \textbf{Steps} & \textbf{Lat. (ms)} & \textbf{Highlights (Sec.~\ref{sec:experimental results}-B)} \\
\midrule

\multirow{4}{*}{\textbf{Channel Extrapolation}} 
& Proposed Foundation Model      & 44.19 & 168.59 & 50 & 51.02 & \textbf{Robust and superior NMSE} \\
& CDDIM (U-Net)          & 3.51  & 13.40  & 50 & 29.88 & Up to \textbf{6 dB} worse than proposed \\
& LWM                    & 44.19 & 168.59 & 1  & 0.65   & Up to \textbf{18 dB} worse than proposed \\
& CGAN                   & 1.56  & 5.94   & 1  & 1.13    & Up to \textbf{13 dB} worse than proposed \\
\midrule

\multirow{3}{*}{\textbf{Channel Estimation}} 
& Proposed Foundation + CE Head   & 47.32 & 180.64 & 50 & 53.54  & \textbf{Highest robustness at low SNRs and $\rho_s$} \\
& LWM + CE Head          & 47.32 & 180.64 & 1 & 0.94  & Moderate gap due to inferior prior representation \\
& ChannelNet             & 0.68  & 2.60   & 1  & 0.70    & Massive gap, especially at non-ideal SNRs or $\rho_s$ \\
\midrule

\multirow{2}{*}{\textbf{User Classification}} 
& Proposed Foundation + CLS Head  & 54.44 & 207.81 & 5  & 0.76   & \textbf{Peak accuracy of $\mathbf{95.87\%}$} \\
& LWM + CLS Head         & 54.44 & 207.81 & 1  & 0.76   & Peak accuracy of 90.73\% \\
\midrule

\multirow{3}{*}{\textbf{User Localization}} 
& Proposed Foundation + LOC Head  & 55.50 & 211.85 & 5  & 1.37  & \textbf{Distance MAE below 2 m} \\
& LWM + LOC Head         & 55.50 & 211.85 & 1  & 1.28   & Higher localization variance \\
& OMP                    & --    & --     & -- & 47.21  & Higher latency and codebook mismatch \\
\bottomrule
\end{tabular*}

\vspace{4pt}
\parbox{\linewidth}{\scriptsize
\textit{Note:} ``Steps'' denotes the number of reverse sampling steps for generative models, and 1 for one-shot baselines. Lat. stands for the average latency.
}
\end{table*}

\subsubsection{Ablation Studies}
To validate the design of the proposed algorithm, we conduct ablation studies on four key hyperparameters and compare the performance summarized in Table~\ref{tab:ablation_all}. Specifically, evaluating the reverse steps $\Delta_t$ in the utilized diffusion scheme reveals a clear trade-off between generative quality and inference latency. The proposed 50-step setting with $\Delta_t=20$ strikes an optimal balance, accelerating inference by nearly $5\times$ compared with the 200-step baseline while maintaining a highly competitive NMSE of -30.08 dB. Alongside this, increasing the Transformer backbone depth $N_{\mathrm{blk}}$ directly enhances representation capability. Therefore, the 10-block configuration is selected as a practical trade-off, since increasing the depth to 12 blocks brings only a modest NMSE improvement while increasing the model size. For downstream adaptation, the pair of frozen and unfrozen fine-tuning blocks $(B,L)$ is critical to effectiveness. Our $(7,3)$ configuration unfreezing the top 3 blocks peaks at the highest accuracy of 95.50\%, effectively avoiding both the severe catastrophic forgetting seen in the $(1,9)$ setting and the underfitting of the single-block $(9,1)$ tuning. Finally, rather than relying on single-timestep features or a 3-timestep fusion, our $5$-timestep strategy seamlessly aggregates the multi-scale features distributed across the diffusion process, ultimately achieving the highest downstream classification accuracy.

\subsection{Computational Overhead Analysis}

Finally, we summarize the computational overhead in Table~\ref{tab:overhead_comparison}.
From the scalability perspective, the parameter cost of our framework is mainly concentrated in the shared networks, i.e., the foundation model ($44.19$\,M) and the multi-time fusion module ($10.25$\,M), which are reused across different downstream tasks to extract universal channel representations. By contrast, the task-specific heads are lightweight, requiring only $3.13$\,M, $0.003$\,M, and $1.06$\,M parameters for channel estimation, user classification, and user localization, respectively. Therefore, adapting the framework to a new task only requires attaching a compact head and performing lightweight fine-tuning on limited task-specific data, rather than designing and training a dedicated model from scratch.

From the latency perspective, the main overhead of the proposed method comes from the iterative reverse sampling process. For channel acquisition tasks, we use 50 reverse sampling steps to improve reconstruction fidelity and robustness, resulting in latencies of 51.02~ms and 53.54~ms for channel extrapolation and channel estimation, respectively. Although slower than one-shot baselines, this iterative design enables more accurate and robust CSI acquisition, with clear NMSE gains under severe masking and low-SNR conditions.

For downstream exploitation tasks, the latency becomes more favorable. As shown in Table~\ref{tab:overhead_comparison}, using only 5 selected diffusion steps is sufficient for user classification and localization, resulting in latencies of 0.76~ms and 1.37~ms, respectively. For these tasks, the latency is mainly dominated by feature extraction and the fusion/head modules, while the additional cost of using a small number of diffusion steps is relatively limited. These latencies remain low and are accompanied by clear performance gains.

For user localization, although OMP requires no trainable parameters, its accuracy is limited by the use of a fixed, discretized near-field codebook, which cannot precisely match or adapt to the actual channel environment. This limitation is further aggravated by the off-grid effect, especially in distance estimation. Improving the precision would require a denser codebook, which would in turn increase the computational cost. Specifically, the complexity of OMP scales as $\mathcal{O}(K_{\mathrm{iter}}MD + K_{\mathrm{iter}}^{2}M + K_{\mathrm{iter}}^{3})$, where $K_{\mathrm{iter}}$, $M$, and $D$ denote the number of OMP iterations, BS antennas, and codebook size, respectively. In contrast, our learning-based localization head avoids codebook matching and achieves both lower latency and higher 3D positioning accuracy.

Overall, the proposed framework offers a reasonable trade-off among scalability, latency, accuracy, and task generality. Its main practical benefit is that it enables high-fidelity CSI acquisition and effective downstream transfer within a unified generative foundation model, while requiring only lightweight task-specific adaptation for new applications.

\section{Conclusion}
\label{sec:Conclusion}

In this paper, we addressed the challenges of dataset scarcity and channel modeling for low-altitude 3D and XL-MIMO systems by introducing \textbf{LAETwin-XL}, a DT-based toolchain and dataset. Based on this dataset, we further developed a CDDIM-based generative foundation model for learning transferable XL-MIMO channel representations from severely incomplete channel observations, together with a downstream fine-tuning paradigm that supports efficient fine-tuning for diverse wireless tasks using only lightweight task-specific heads. Experimental results demonstrated that the proposed framework not only achieves strong zero-shot generalization for antenna-domain channel extrapolation, but also enables efficient adaptation to various downstream tasks. Overall, LAETwin-XL and the proposed generative framework provide an effective data resource and a powerful methodology for AI-assisted 6G wireless communication research, specifically tailored for low-altitude XL-MIMO systems. For future work, as practical XL-MIMO testbeds become available, we will further validate the learned generative priors using real-world measurements. Moreover, we will also explore accelerated sampling strategies to further reduce the computational overhead and latency for hardware deployment.

 \small\bibliographystyle{./bibliography/IEEEtran}
 \bibliography{ref_modified}

\vspace{12pt}
\color{red}

\end{document}